\documentclass[useAMS,usenatbib,graphicx]{mn2e}

\usepackage{graphicx}

\title[MHD simulations of feedback in a sheet-like cloud]{Magnetohydrodynamic simulations of 
 mechanical stellar feedback in a sheet-like molecular cloud}
\author[C. J. Wareing et al.]{C. J. Wareing$^{1}$\thanks{E-mail:
C.J.Wareing@leeds.ac.uk}, J. M. Pittard$^{1}$ and S. A. E. G. Falle$^{2}$\\
$^{1}$School of Physics and Astronomy, University of Leeds, Leeds, LS2 9JT, U.K.\\
$^{2}$School of Mathematics, University of Leeds, Leeds, LS2 9JT, U.K.}
\begin{document}

\date{Accepted 2016 November 16. Received 2016 November 16; in original form 2016 May 16}

\pagerange{\pageref{firstpage}--\pageref{lastpage}} \pubyear{2002}

\maketitle

\label{firstpage}

\begin{abstract}
We have used the AMR hydrodynamic code, MG, to perform 3D
magnetohydrodynamic simulations with self-gravity of stellar feedback in a sheet-like
molecular cloud formed through the action of the thermal
instability. We simulate the interaction of the mechanical energy
input from a 15\,M$_\odot$ star and a 40\,M$_\odot$ star into a
100\,pc-diameter 17000\,M$_\odot$ cloud with a corrugated
sheet morphology that in projection appears filamentary.  The
stellar winds are introduced using appropriate Geneva
stellar evolution models. In the 15\,M$_\odot$ star case, the wind
forms a narrow bipolar cavity with minimal effect on the parent
cloud. In the 40\,M$_\odot$ star case, the more powerful stellar wind
creates a large cylindrical cavity through the centre of the cloud.
After 12.5\,Myrs and 4.97\,Myrs respectively, the massive stars
explode as supernovae (SNe). In the 15\,M$_\odot$ star case, the SN
material and energy is primarily deposited into the molecular cloud
surroundings over $\sim 10^{5}$ years before the SN remnant escapes
the cloud. In the 40\,M$_\odot$ star case, a significant fraction of the SN
material and energy rapidly escapes the molecular cloud along the
wind cavity in a few tens of kiloyears. Both SN events compress the
molecular cloud material around them to higher densities (so may
trigger further star formation), and strengthen the magnetic field,
typically by factors of 2-3 but up to a factor of 10. Our simulations 
are relevant to observations of bubbles in flattened ring-like
molecular clouds and bipolar HII regions.
\end{abstract}


\section{Introduction}

Stellar feedback, through both winds and supernovae (SNe), is
recognized as having significant influence on galactic and
extra-galactic scales. Winds and SNe not only chemically enrich the
interstellar medium (ISM), but also drive structural evolution and
motion within it. Massive stars embedded within molecular clouds most
likely affect star formation, with the ability to both inhibit further
star formation, as their winds and ionizing radiation disperse the
molecular gas, and in some cases trigger new star formation
\citep{koenig12} and even new cluster formation
\citep{beuther08,gray11}.
Winds are particularly important to consider as they
can introduce similar amounts of mass and energy as SNe and 
strongly affect the environment into which SNe explode.

Stellar wind feedback into an inhomogeneous environment has been
studied by \cite{harper09}.  They postulate that the non-uniform
surrounding medium causes gaps in the swept-up shell surrounding the
wind-blown bubble where some of the high-pressure gas in the bubble
interior can leak out. \cite{lopez11} supported this scenario
concluding that such leakage may be occurring within 30 Doradus. They
also conclude that direct stellar radiation pressure dominates the
interior dynamics, but this claim has proved far more controversial,
and other works are in favour of the thermal pressure of hot X-ray
emitting plasma shaping the large-scale structure and dynamics in 30
Doradus \citep[e.g.][]{pellegrini11}. More recently, \cite{lopez14}
concluded that the warm ionised gas pressure dominates in HII regions
on scales of $10-100$\,pc.

The possibility that the pressure exerted by stellar radiation may be
dynamically important in massive young stellar clusters has also
received much attention in recent years, with \cite{krumholz09},
\cite{fall10} and \cite{murray10} all arguing that radiation pressure
is the dominant feedback mechanism. However, these works disagree on
the net momentum coupling between the radiation field and the gas,
partially because this depends on the degree of inhomogeneity of the
gas and the effect that this has on the radiation field
\cite[e.g.][]{krumholz12}. Complementary work on the ionised gas
pressure has shown that ionization feedback into a highly
inhomogeneous medium is not very effective at high cluster masses
\citep{dale11}, but becomes more so at lower masses
\citep{dale12,walch12}.

Recent work has also explored the effect of supernovae on the
multi-phase ISM and molecular clouds
\citep{gatto15,walch15,walch15b,giri15,kortgenphd,kortgen16}.
\cite{giri15} employed 3D MHD simulations of the ISM in a vertically
stratified box including self-gravity, an external potential due to
the stellar component of the galactic disc, and stellar feedback in the
form of an interstellar radiation field and SNe. They also used a
sophisticated cooling prescription based on full tracking of a
chemical network following the abundances of H$^{+}$, H, H$_2$,
C$^{+}$ and CO, taking into account shielding in a consistent
fashion. They varied the SN feedback with different rates, clustering
and positioning in the disc. Positioning, either randomly or at peak
density locations, had a major impact on the dynamics. Only for random
SN positions is the energy injected in sufficiently low-density
environments as to reduce energy losses and enhance the kinetic
coupling of the SNe with the gas, leading to more realistic velocity
dispersions and strong outflows. SNe placed in density peaks do not
drive any noticeable outflow. However, the effect of the SN-preceding
stellar wind was not accounted for and very recent
simulations \citep{simpson16} accounting for the production of cosmic rays in supernova
events has shown that they are able to drive outflows from SNe placed
at density peaks, with similar
mass loading as obtained from random placement of SNe with no
cosmic rays. 

\cite{offner15} used MHD simulations to model winds launching within
turbulent molecular clouds and explored the impact of wind properties
on cloud morphology and turbulence.  They used the AMR code ORION to
model an isothermal molecular cloud.
In 7 out of 9 cases they use a $\beta$ of 0.1 -
a magnetic pressure $10\times$ greater than the thermal pressure and
hence a magnetic field that strongly influences the evolution. They
introduce stellar winds with constant mass-loss rates and constant velocities.
They found that the winds do not produce clear features in turbulent
statistics such as the Fourier spectra of density and momentum, but do
impact on the Fourier velocity spectrum.  They showed that stellar
mass-loss rates for individual stars must be greater than
10$^{-7}$\,M$_\odot$\,yr$^{-1}$, similar to those estimated from
observations, in order to reproduce shell properties. 

K\"ortgen, in his recent PhD work \citep{kortgenphd} and resulting
publication \citep{kortgen16}, performed MHD simulations of converging
turbulent (Mach $\sim$ 2) warm neutral medium flows in order to form
magnetised and turbulent molecular clouds. SN feedback was then
introduced into these clouds, through the form of thermal and kinetic
energy according to E$_{SN}$=10$^{51}$\,erg with
E$_{th}$=0.65\,E$_{SN}$ and E$_{kin}$=0.35\,E$_{SN}$ in a spherical
control volume of radius 0.06\,pc. They found that SN feedback alone
failed to disrupt the entire molecular cloud, but was able to disperse
small-sized ($\sim$10\,pc) regions on timescales of less than
1\,Myr. Efficient radiative cooling of the SN remnant as well as
strong compression of the surrounding gas result in non-persistent
energy and momentum input from the SNe.  Multiple clustered SNe with
short intervals (at a rate much higher than typical IMF calculations)
are able to combine to form large hot bubbles that disperse larger
regions of the parental cloud, and give an {\em upper} limit on the
efficiency of SN feedback given the assumptions in the model.
\cite{kim15} also analysed SNe exploding in a two-phase ISM \citep[see
also][]{martizzi15} with no feedback prior to the SNe and found the
net momentum input is nearly independent of the environment. However,
this conclusion may well change if pre-SN (wind and ionizing
radiation) feedback is considered.

Simulations of momentum-driven or isothermal winds
\citep{dale14,dale15,offner15} give a lower limit to their
impact. Pre-SN feedback has been found to enhance the impact of SNe
\citep{rogers13,fierlinger16}, and, in whole galaxy models, clears
dense gas from star forming regions, reducing the star formation rate
\citep[e.g.,][]{agertz13}. However, whole galaxy models remain very
sensitive to assumptions in the feedback scheme.
 
In recent work by our group \citep{rogers13,rogers14}, we examined the
extent to which the mechanical energy input from a cluster of massive
stars is confined by and shapes the local environment, focusing on
feedback due to SNe and the preceding stellar winds. 3D hydrodynamic
models simulated the interaction of the mechanical energy from three
massive stars with a giant molecular cloud clump containing
3240\,M$_\odot$ of material within a 4\,pc radius. The stars were
evolved through 3 idealized evolutionary phases: O-star main sequence (MS),
red supergiant (RSG), and Wolf-Rayet (WR), before exploding as
SNe. The clump structure was based upon the simulations of
\cite{vazquez08} of turbulent and clumpy molecular clouds.
The combined cluster wind from the three O-stars was found to blow
out of the inhomogeneous molecular clump along paths of least
resistance. Hot, high-speed gas flows away from the cluster through
these low-density channels, into which denser clump material is
entrained. These channels are directly related to the initial cloud
structure. Molecular material is gradually removed by the cluster wind
during which mass-loading factors in excess of several 100 were
obtained, but this process is relatively slow and a substantial amount
of molecular material remains near the cluster when the first SN
occurs.
However, because this material has a small volume filling fraction,
the environment is highly porous to the transport of wind and SN
energy, and the majority of the wind energy and essentially all of the
SN energy escaped to the wider surroundings. Nevertheless, the winds
appeared to be better at removing molecular material from the cluster
environment, despite injecting less energy then the SNe. The molecular
material was found to be almost completely dispersed and destroyed
after 6\,Myr. This work demonstrated the complexity of the interaction
of a cluster wind with an inhomogeneous environment, and is far
removed from simple spherically symmetric models. 

A key conclusion of \cite{rogers13} is that massive stars that form at
{\it high density} locations {\em are} able to disperse surrounding
material through the action of their winds, aiding the
relatively unhindered escape of SN energy. Therefore, the effect of the
preceding stellar wind phase should be considered when calculating the
impact of SN explosions on the surrounding gaseous environment.

In our previous work the simulated results were strongly linked to the
initial clump structure and did not include magnetic fields, realistic
stellar wind evolution or self-gravity. A parameterised
heating/cooling rate per unit volume was also employed.  In the work
presented here we make several improvements in our exploration of the
effect of stellar feedback on molecular clouds. Firstly, we use a
self-consistent cloud structure, which has formed naturally under the
influence of the thermal instability, magnetic fields and self-gravity
\citep[hereafter Paper I]{wareing16}. In our previous paper feedback
occurred into a roughly spherical, but inhomogeneous, region. Here we
explore the other extreme, where the stellar feedback occurs into a
sheet-like structure, as formed when the initial plasma beta is
unity\footnote{As far as we are aware this is the first time that a
  sheet-like cloud morphology has been shown to form through the
  action of the thermal instability \citep[colliding flows
  are another mechanism - see, e.g.,][]{heitsch05,vazquez06}.}. Such
flattened cloud morphologies have been inferred in recent work by
\cite{beaumont10}. We also examine feedback from a single star (of
mass 15\,M$_\odot$ and 40\,M$_\odot$), employing realistic Geneva
stellar evolution tracks for the time-varying wind properties, rather
than the multiple stars and simplified wind evolution considered by
\cite{rogers13}.  Finally, we use a more accurate heating and cooling
prescription (albeit leaving photoionisation and the full tracking of
chemical species and their interplay to a future work).

In the next section, we present our numerical method and define
the initial conditions used in our model. In Section \ref{results} we
present and discuss our results. In Section \ref{analysis}
we consider the global energy, density, phase and mixing behaviour. 
In Section \ref{comparisons} we
further analyse those results with comparison to previous work and relevant
recent observational results. We summarise and conclude the work in
Section \ref{conclusions}.

\section{Numerical Methods and initial conditions}\label{numerical}

\subsection{Numerical methods}

We present 3D, magneto-hydrodynamical (MHD) simulations of stellar
feedback with self-gravity using the established astrophysical code MG
\citep{falle91} as recently used in Paper I. The code employs an
upwind, conservative shock-capturing scheme and is able to employ
multiple processors through parallelisation with the message passing
interface (MPI) library. MG uses piece-wise linear cell interpolation
to solve the Eulerian equations of hydrodynamics. The Riemann problem
is solved at cell interfaces to obtain the conserved fluxes for the
time update. Integration in time proceeds according to a second-order
accurate Godunov method \citep{godunov59}. A Kurganov Tadmor Riemann
solver is used in this work \citep{kurg00}. Self-gravity is computed
using a full-approximation multigrid to solve the Poisson
equation. The magnetic field components are cell-centered and the
$\nabla \cdot B$ criterion is enforced using divergence cleaning
\citep{dedner02}.

The adaptive mesh refinement (AMR) method \citep{falle05} employs an
unstructured grid approach. By default, the two coarsest levels (G0
and G1) cover the whole computational domain; finer grids need not do
so. Refinement or derefinement is based on error. Where there are
steep gradients of variable magnitudes such as at filaments, flow
boundaries or discontinuities, this automated meshing strategy allows
the mesh to be more refined than in more uniform areas.  Each level is
generated from its predecessor by doubling the number of computational
grid cells in each spatial direction. This technique enables the
generation of fine grids in regions of high spatial and temporal
variation, and conversely, relatively coarse grids where the flow
field is numerically smooth. Defragmentation of the AMR grid in
hardware memory is performed at every time-step, gaining further speed
improvements for negligible cost through reallocation of cells into
consecutive memory locations.  The simulations presented below
employed 8 levels of AMR at a minimum, with resolution tests to 10
levels of AMR. Physical domain sizes and hence physical resolutions
are as detailed below.

\subsection{Heating and cooling processes}

Care has been taken to continue our implementation of realistic
heating and cooling as used in Paper I, as it is the balance of these
processes that formed the initial molecular cloud condition. In the
ISM, heating as defined by the coefficient $\Gamma$, varies with
increasing density as the starlight, soft X-ray and cosmic ray flux
are attenuated by the high column density associated with dense
clouds. Because the exact form of the attenuation depends on details
which remain uncertain (e.g. the size and abundance of PAHs), the
heating rate at T $\leq$ 10$^4$\,K is similarly uncertain. In this
work, following on from the filament formation work, we have assumed
that $\Gamma = 2\times10^{-26}$ erg\,s$^{-1}$ (independent of density
or temperature). For the low-temperature cooling ($\leq10^4$ K), we
have followed the detailed prescription of \cite{koyama00}, fitted by
\cite{koyama02}, and corrected according to \cite{vazquez07}, namely
\begin{equation}
\begin{array}{ll}
\displaystyle{\frac{{\Lambda (T)}}{\Gamma } =} & \displaystyle{{10^7}\exp \left( {\frac{{ - 1.184 \times {{10}^5}}}{{T + 1000}}} \right)}\\
&\\
 & \displaystyle{+ 1.4 \times {10^{ - 2}}\sqrt T \exp \left( {\frac{{ - 92}}{T}} \right)}.
\end{array}
\end{equation}
At higher temperatures we have followed the prescription of
\cite{gnat12} who used CLOUDY 10.00, enabling us to define cooling
rates over the temperature range from 10\,K to 10$^8$\,K.  This has
been implemented into MG as a lookup table for efficient computation.

\subsection{Initial conditions}

\begin{figure*}
\centering
\includegraphics[width=150mm]{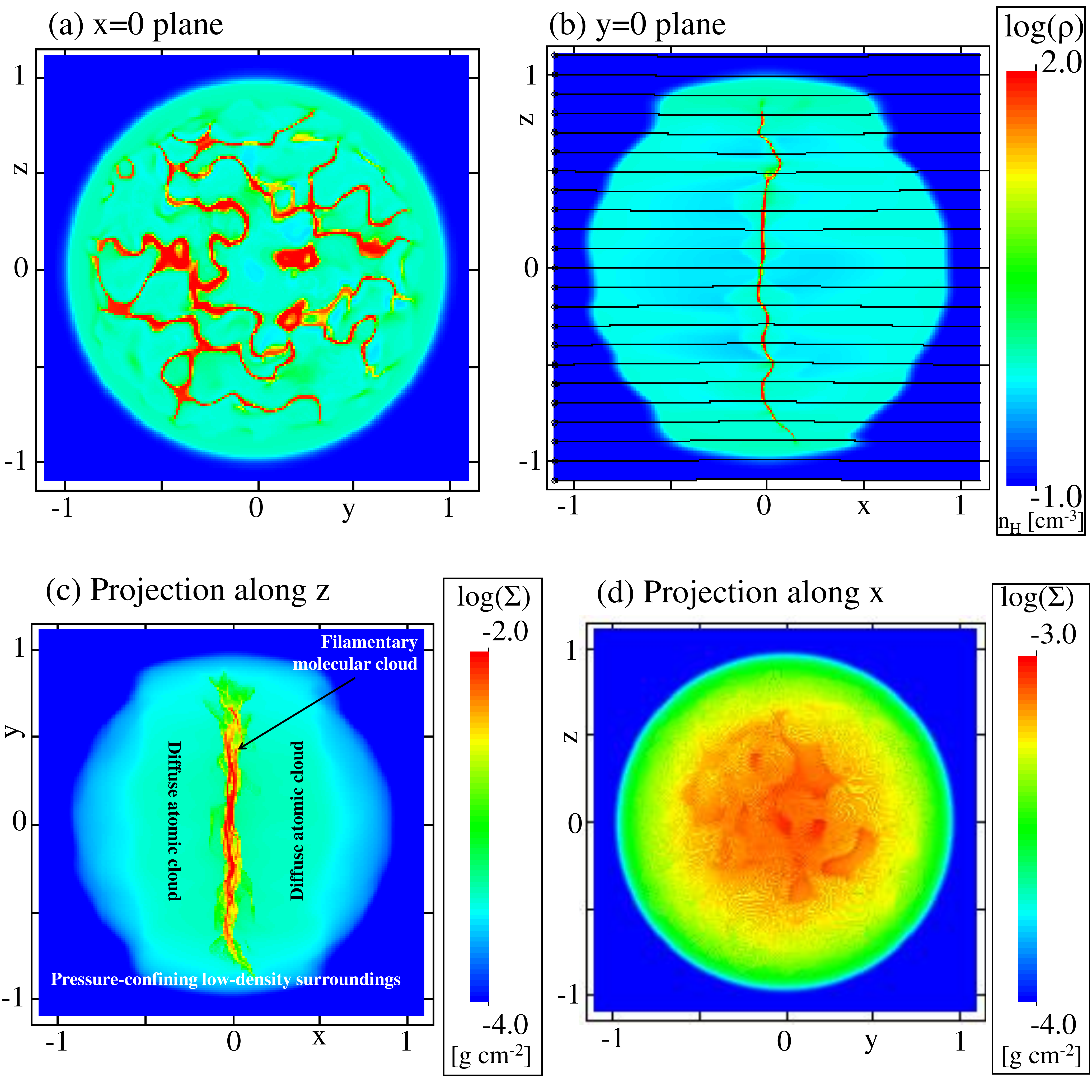}
\caption{Initial condition. Snapshot of a $\beta=1.0$ filamentary cloud after 38.8\,Myrs 
of evolution used as the initial condition in this work. Shown are
logarithm of mass density on planes (a) x=0 and (b) y=0 and projected column density
perpendicular to the field direction in (c) and parallel to the field direction in (d). 
Length is scaled in units of 50\,pc. Field lines are shown where appropriate.} 
\label{fig1}
\end{figure*}

Paper I focused on the study of filamentary molecular cloud formation
under the influence of the thermal instability
\citep{parker53,field65}. In that work, we examined the evolution of
diffuse clouds varying $\beta$, the ratio of thermal pressure to
magnetic pressure.  We examined three cases, $\beta$ = 0.1, 1.0 and
$\infty$ (equivalent to the hydrodynamic case of zero magnetic
field). Following Paper I, the initial condition consists of a
stationary cloud of radius 50\,pc with a number density of hydrogen
throughout the cloud of n$_H$ = 1.1 cm$^{-3}$. In the
cloud, 10\% density variations about the uniform initial density were
introduced, giving the cloud a total mass of
$\sim$17,000\,M$_\odot$. The pressure was set according to the local
density and thermal equilibrium between heating and cooling
prescriptions at P$_{eq}$/k = $4700\pm300$ K\,cm$^{-3}$, resulting
in an initial temperature T$_{eq}$ = $4300\pm700$\,K (an unstable part 
of the equilibrium curve - for more details see Paper I). The pressure of
the lower density (n$_H$ = 0.1 cm$^{-3}$) surroundings was set 
equal to that of the unperturbed cloud.

The domain was initially threaded with uniform $B$-field along the $x$
direction, i.e. \textbf{B} = $B_0$ $\hat{\textbf{I}}_x$. For the
moderate field case of $\beta = 1.0$, the initial cloud condition was
pressure equality, i.e. the thermal pressure of the initial cloud with
n$_H$ = 1.1 cm$^{-3}$ led to $B_0$ = 1.15\,$\mu$G. This is a typical
Galactic field strength at an inner ($\sim$ 4\,kpc) location.  Evolved
molecular clouds have been noted to have magnetic field strengths over
10$\times$ greater than these, but it is not clear how these field
strengths have been generated. Having shown in Paper I that filament
formation under the action of the thermal instability cannot intensify
the magnetic field, we also examine in this work whether stellar
feedback generating super-sonic, super-Alfv{\'e}nic motions can
intensify the field to such levels.

The computational domain consists of a 150\,pc$^3$ box with free-flow
boundary conditions (non-periodic for self-gravity)
and AMR level G0 containing 4$^3$ cells. Eight
levels of AMR mean an effective resolution of 512$^3$ cells, although
G5 with 128$^3$ cells is the finest fully populated level (rather than
the MG default of G1, in order to resolve the action of the thermal
instability). Such a large number of AMR levels is employed in order
to efficiently compute the self-gravity on the coarsest levels and
also fully resolve the structures formed in the molecular cloud. The
finest physical resolution is 0.293\,pc.
The large number of AMR levels is computationally
costly. Each 3D MHD simulation with stellar feedback and self-gravity 
presented here with 8 AMR levels took approximately 60,000 CPUhours 
($\sim$10-12 48-hour cycles on 128 cores of the MARC1 high performance computing facility at 
Leeds), so 120,000 CPUhours for the two runs presented. Supporting 
investigations including a parameter exploration of the initial condition 
(see Paper I), high resolution tests, stellar feedback tests \& 
tests with other stellar masses etc. mean the total computational cost 
of the work presented in this paper approaches 200,000 CPUhours. A
large-scale parameter exploration is thus prohibitively expensive. Regarding
the choice of number of AMR levels used for the base case, each
further level of AMR introduces a computational-cost-multiplier of between
3 and 4 for this model, thus making complete higher resolution simulations prohibitively
expensive at this time.

The influence of the thermal instability causes the cloud to evolve
into multiple clumps without any magnetic field and into a filamentary
structure under the influence of magnetic field. Specifically, with
$\beta=\infty$, molecular clumps formed across the diffuse cloud. In
the other two cases, molecular filaments formed across the cloud,
eventually forming intersecting bundles of filaments and corrugated
sheets. Seen in projection, the filaments closely resemble
those observed. Densities, column densities, widths, velocity
dispersions and separations all resemble derived properties of
observed filaments. Self-gravity accelerates the
contraction of the cloud and makes the cloud more filamentary.
Besides exploring the importance of the thermal
instability in molecular cloud evolution, the secondary aim of Paper I
was to provide a more realistic initial condition for this work, by
including more accurate heating and cooling, the effect of thermal
instability, self-gravity and magnetic fields, as compared to our
previous feedback studies \citep{rogers13,rogers14}.

\begin{figure*}
\centering
\includegraphics[width=175mm]{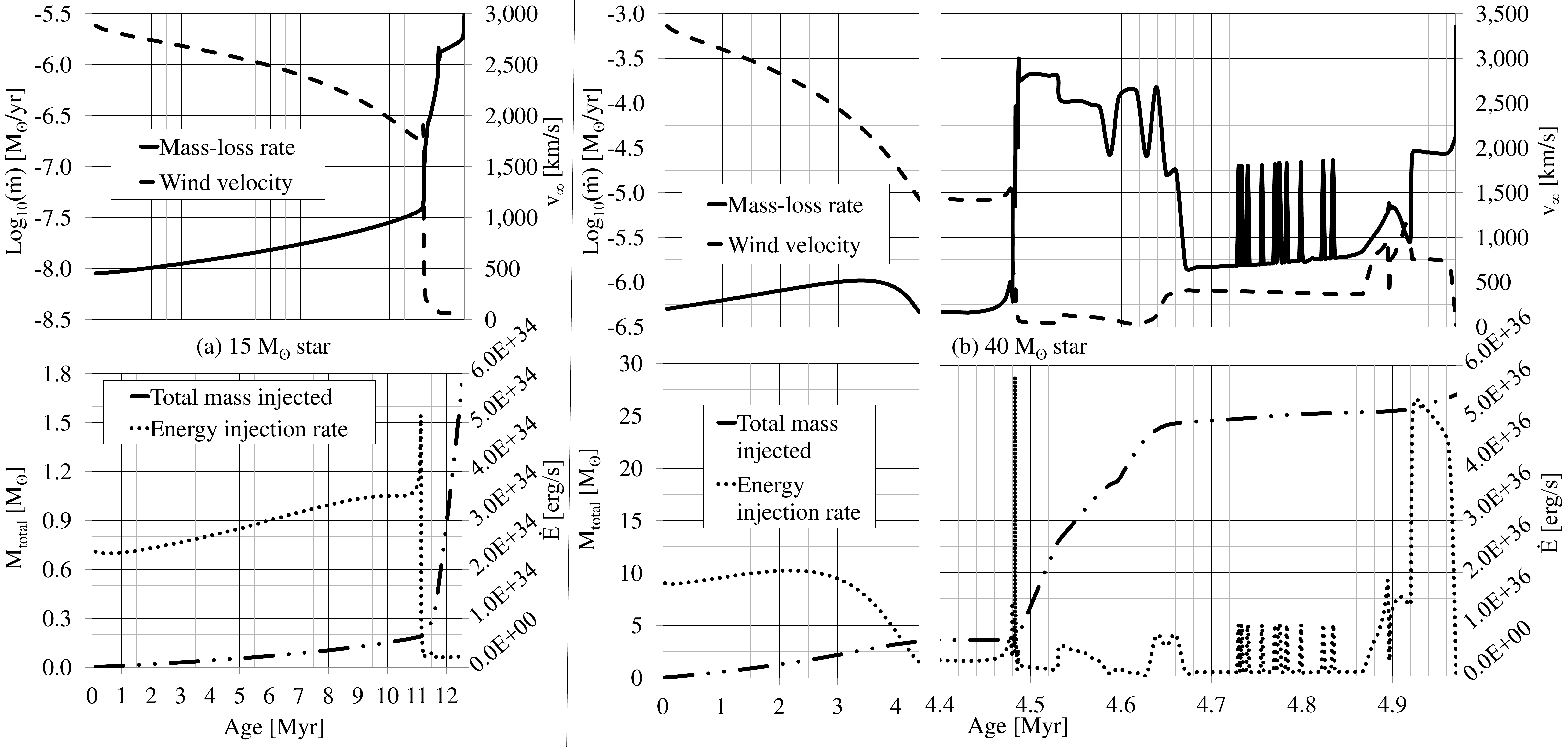}
\caption{Stellar evolution tracks \citep{vink00,vink01,ekstrom12} for 
the (a) 15\,M$_\odot$ star and the (b) 40\,M$_\odot$ star, showing 
mass-loss rate and wind velocity on the upper graphs, and energy 
injection rate and total injected mass on the lower graphs.} 
\label{fig2}
\end{figure*}

In this work, we take as our initial condition a repeat of a
$\beta=1.0$ filamentary cloud simulation following the method of Paper
I. Different random seeds results in a simulation that is
qualitatively the same as the $\beta=1.0$ case in Paper I, but
quantitatively different.  After 32.9\,Myrs of evolution, densities in
the filamentary structures formed in this new simulation have reached
100\,cm$^{-3}$ - the density threshold often used for injection of
stars in similar simulation work \citep[e.g.,][]{fogerty16}. At a
density of 100\,cm$^{-3}$, the free-fall time due to gravity is
5.89\,Myrs. Therefore, we choose to introduce stars into our
simulation after 38.8\,Myrs of evolution. We show snapshots of this
initial condition in Figure \ref{fig1} and refer the interested reader
to Paper I for a full description of the evolutionary process that led
to the formation of this cloud. It is important to note though that
this time-scale should not be considered as the `age' of the parent
molecular cloud - this is the length of time required to go from
a diffuse cloud with an average density of n$_H$ = 1.1 cm$^{-3}$ to
a structured molecular cloud where feedback can be introduced. 
Grid levels G6 and G7 are 18\% and 10\% populated
respectively at this simulation time, leading to an initial total of
$18.5\times10^6$ grid cells. 

We consider two scenarios, each employing this initial condition, in
order to examine the effect of stellar feedback.

\subsubsection{Scenario 1 - a 15\,M$_{\odot}$ star}

In this scenario we introduce a 15\,M$_\odot$ star at the position
(x,y,z) = (-0.025, 0.0, 0.0125) where the coordinates are given in
scaled code units.  This location has been selected as the highest
density location in the sheet, closest to the centre of the
domain. Further, enough mass is present in a spherical volume with a
5-cell radius centred at this point, that can form a 15\,M$_\odot$
star assuming $\sim$100\% material conversion from cloud to star. For
this first investigation, the mass is removed at the switch on of the stellar 
wind, t$_{wind}$=0. Whilst this may seem inconsistent, if the mass is
left in the injection region, the stellar wind rapidly and
unrealistically cools and hence the feedback effects are significantly
underestimated. 
The location of the star remains constant throughout this work. In future
work, we plan to convert the mass into a `star' particle following the
method in MG of \cite{vanloo13} and \cite{vanloo15}.  The star can
then move through the domain whilst feeding back into the domain
through winds and SNe and remain consistent with self-gravity in the
simulation. The stable position of the molecular cloud does not require
particles for these simulations. 

For the stellar evolution, a 15\,M$_{\odot}$ non-rotating Geneva stellar
evolution model calculated by \cite{ekstrom12} is used in order to provide a
realistic mass-loss rate over the lifetime of the star. We use the
mass, temperature, luminosity and metallicity together with the
prescriptions of \cite{vink00,vink01} in order to calculate the wind
velocity. \cite{ekstrom12} describe lower mass star tracks, but the
minimum stellar mass that can be used with the \cite{vink00,vink01}
method is 15\,M$_{\odot}$. The calculated mass-loss rate and wind
velocity are shown in Fig.~\ref{fig2}(a). Also shown in
Fig.~\ref{fig2}(a) are the energy injection rate and total injected
mass. The stellar wind is introduced as a mass and energy source,
directly using the mass-loss rate as a density source and the kinetic
energy of the wind as a thermalised energy source. No momentum source
is introduced in this work. The wind is introduced over the same
spherical region of the grid with a radius of 5 cells at the finest
grid level that was used to remove the equivalent or greater mass of
molecular material.  An advected scalar, $\alpha_{wind}$, previously
set to zero throughout the grid, is set to 1 in the wind injection
region in order to track the movement and mixing of the wind material.
The total mass and total energy injected by the star prior to
supernova explosion are 1.75\,M$_\odot$ and $1.05\times10^{49}$\,erg
respectively. The wind rapidly expands outside the injection region.

After 12.5\,Myrs, the star explodes as a SN, injecting 10\,M$_\odot$
of stellar material and 10$^{51}$ ergs of energy into the same wind
injection volume. The mass and energy is injected over 500\,yrs,
roughly consistent with the time taken for a remnant to reach the size
of the injection volume. A second advected scalar, $\alpha_{SN}$,
previously set to zero throughout the grid, is set to 1 in the
supernova injection region in order to track the movement and mixing
of the SN material. At this time the wind scalar $\alpha_{wind}$ is
set to zero.

Limited numerical convergence testing has shown the introduction of a
finer grid level accelerates the wind more quickly given the increased
source terms, but that the simulations here capture the large-scale
interaction of the wind and filamentary molecular cloud that we are
interested in. Computationally, even the use of a single further fine
AMR level becomes very expensive and time-consuming,
as discussed in the preceding subsection.

Gravity plays less of a role during the stellar wind phase and the early SN
  phase, due to the comparatively powerful dynamics even in the
  15\,M$_\odot$ star case, but we continue to include it
for consistency within the cloud and to explore the evolution
post-SN when it again plays more of a role. However, 
assessing gravity's relative importance post-SN, as
compared to the flow dynamics induced by the SNR, 
the triggered thermal instability, and 
any numerical effects now present is difficult.

\subsubsection{Scenario 2 - a 40\,M$_{\odot}$ star}

In this scenario we introduce a 40\,M$_\odot$ star using the same
method and position as in scenario 1.  A spherical volume of material
slightly larger than in the previous scenario was removed in order to
conserve mass in introducing the star, but again this was small enough
to be unimportant in terms of the global evolution of the wind-cloud
interaction. A 40\,M$_{\odot}$ non-rotating stellar evolution model calculated by
\cite{ekstrom12} is used in order to provide a realistic mass-loss
rate over the lifetime of the star in the same manner as in the first
scenario. The calculated mass-loss rate and wind velocity are shown in
Fig.~\ref{fig2}(b). Also shown in Fig.~\ref{fig2}(b) are the energy
injection rate and total injected mass. The total mass and total
energy injected by the star prior to supernova explosion are
27.2\,M$_\odot$ and $2.50\times10^{50}$\,erg respectively,
comparable to the material and energy introduced in a SN event. After
4.97\,Myrs, the star explodes as a SN in the same manner as in the
first scenario.

\subsection{Neglected processes and simplifications}

This work represents a step forward in modelling mechanical stellar
feedback in molecular cloud conditions. We have necessarily made a
number of simplifications and approximations, a number of which have
been outlined above but for completeness we consider the rest here.

We neglect the role of carbon monoxide (CO). Without a full treatment
of heating according to column density and shielding to allow the
formation of CO, it is difficult to justify the inclusion of any CO
effects into the cooling curve, although this was done in our previous
work through an amendment to the parameterised low temperature cooling
\citep{rogers13,rogers14}.  In terms of molecular cloud evolution, we
found in tests carried out as part of our previous work
\citep{wareing16} that the increased cooling introduced by CO at lower
temperature allows the clumps and filaments to cool further (and
increase in density) to temperatures on the order of 10-15\,K. With
regard to this work, the tenuous stellar winds and SNe interact with
this material and it is therefore likely that we observe the {\it
  maximum} effect that winds and SNe can have on the filamentary
structures, as higher densities in these structures will lead to
longer ablation times.

We have not included sink particles at  this time, but,  as already
noted, the cloud does not contract further due to 
self-gravity and we plan to adopt this  technique in future work. However, there
is,  as   yet,  no   really  satisfactory   way  of   identifying  the
gravitationally  collapsing  regions  that  can be  turned  into  sink
particles.

Photoionisation and radiative feedback are not included at this
time. Inclusion of such should speed up the destruction of molecular
material. Having said this, photoevaporation may be suppressed in
regions where the ram or thermal pressure of the surrounding medium is
greater than the pressure of the evaporating flow \citep{dyson94}.
Either way, the results of \cite{simpson16} suggest
further study in this area is required as a matter of urgency and we
are developing numerical techniques in order to include these effects.

We also note that dust can dramatically affect the cooling within hot
bubbles if it can be continuously replenished, perhaps by the
evaporation/destruction of dense clumps \citep{everett10} which will
also mass-load the bubble \citep[see e.g.][]{pittard01a,pittard01b}.
The presence of dust will also affect the photoionization rate
throughout the cloud. Clearly the effects of dust warrant further
study.

\section{Results}\label{results}

In this section we present our results. We present both 2D slices
through the computational domains created within MG and 3D contour and
volume visualisations, created using the VisIt software \citep{visit}.
Line profiles on the 2D slices are shown in the Appendix. 
Raw data can be obtained from http://doi.org/10.5518/114.

\subsection{Scenario 1 - a 15\,M$_{\odot}$ star}\label{res-case1}

\begin{figure*}
\centering
\includegraphics[width=155mm]{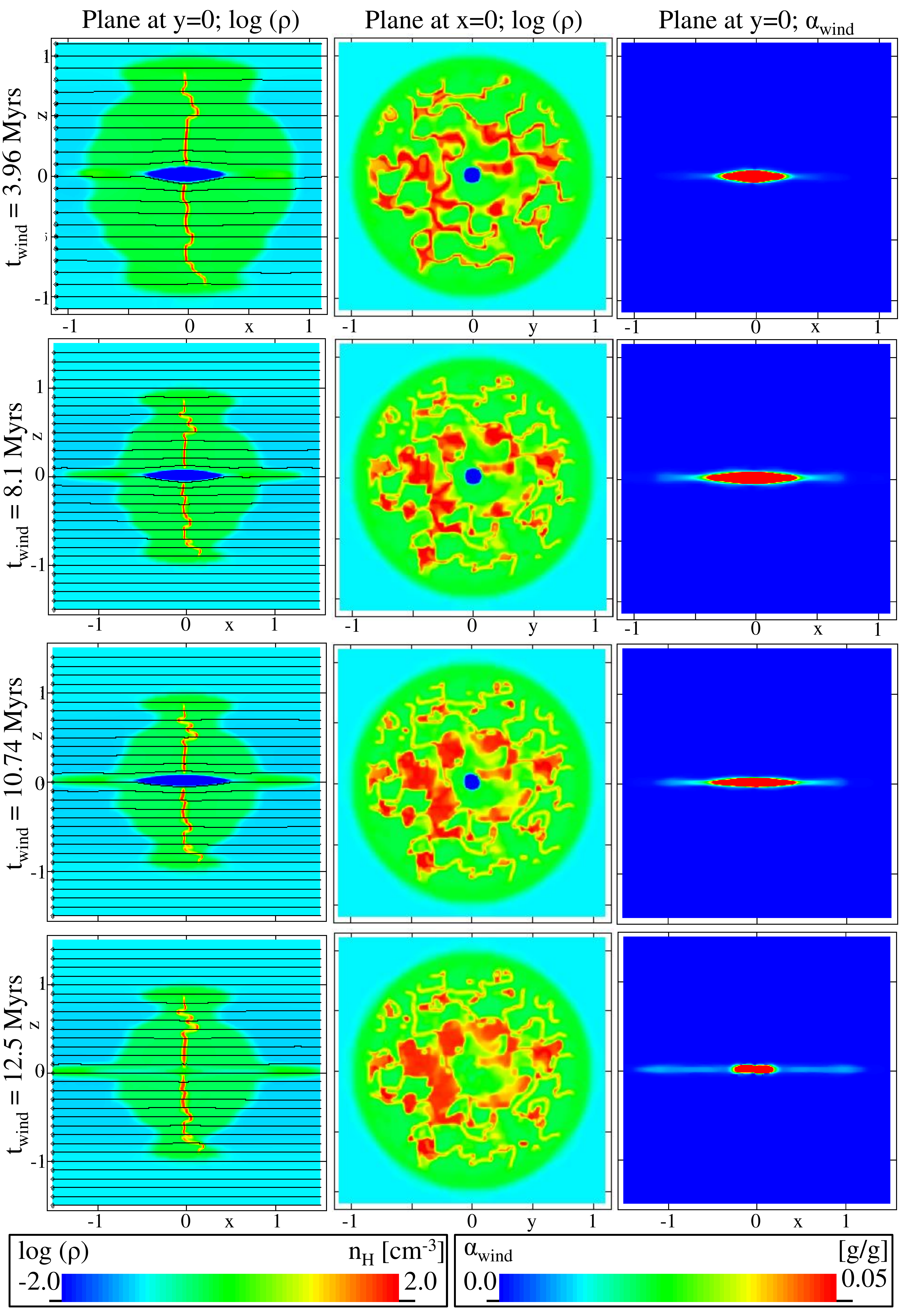}
\caption{Cloud-wind interaction during the lifetime of a 15 M$_\odot$ star. Shown is:
column 1 - the logarithm of mass density on the plane at $y=0$; column 2 - the
logarithm of mass density on the plane at $x=0$; column 3 - the wind scalar on the
plane at $y=0$. Length is scaled in units of 50\,pc. 
For line profiles on these planes, see Figure \ref{figA1}.} 
\label{fig3}
\end{figure*}

\begin{figure*}
\centering
\includegraphics[width=155mm]{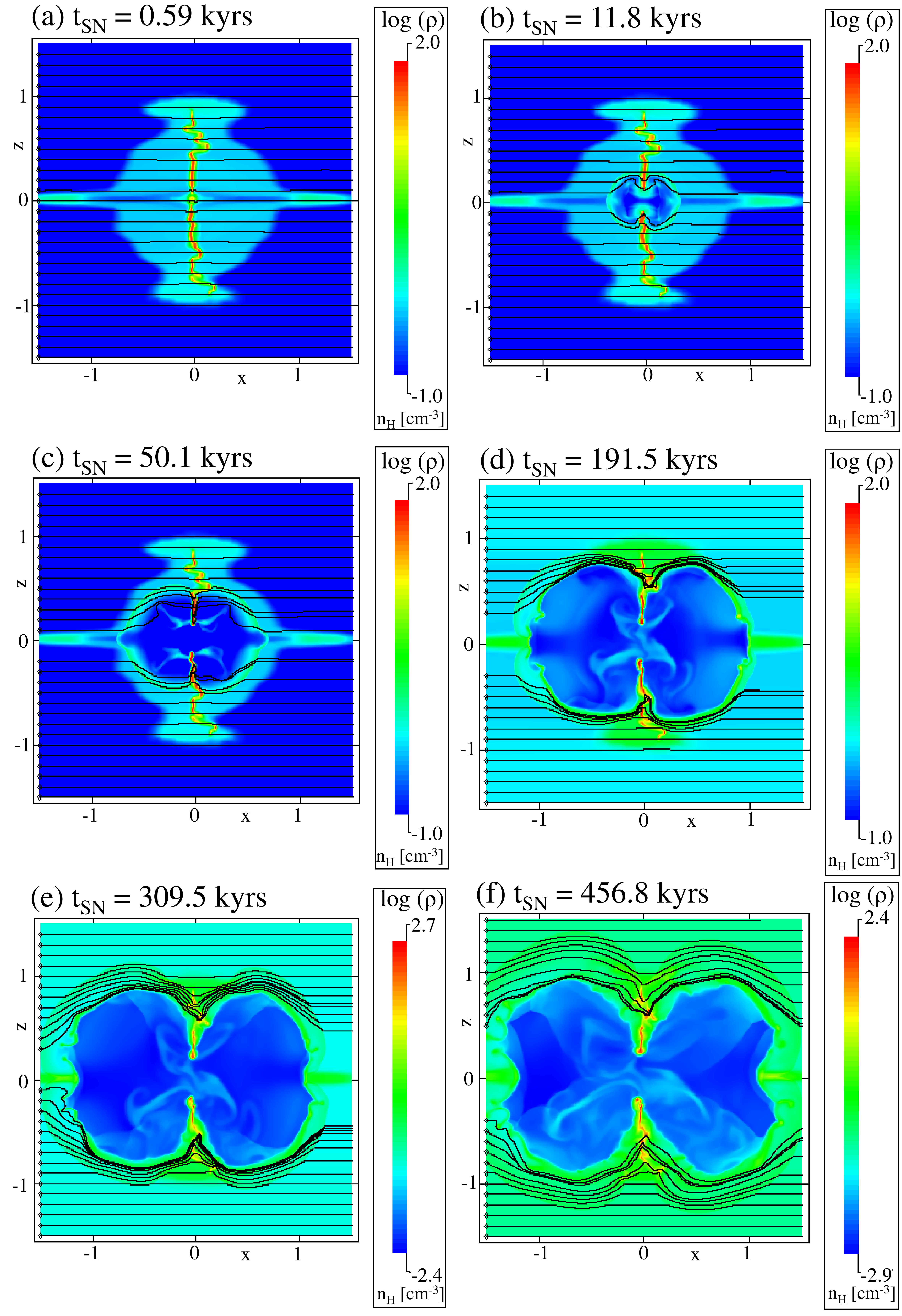}
\caption{SN-wind-cloud interaction during the SN phase of a 15 M$_\odot$ star. Shown 
is the logarithm of mass density on the plane at y=0 at various times through the SN
evolution. Length is scaled in units of 50\,pc. 
For line profiles on these planes, see Figure \ref{figA2}.} 
\label{fig4}
\end{figure*}

In Fig.~\ref{fig3} (see also Fig.~\ref{figA1}) we show density slices through the domain during
the evolution of the 15\,M$_\odot$ star up to the end of its life. The
low mass-loss rate and corresponding low energy injection rate have
minimal effect on the cloud structure, generating only a small bipolar
cavity parallel to the magnetic field at the location of the
star. This creates a hole in the corrugated sheet of filamentary
molecular cloud, but does not emerge from the surrounding diffuse
cloud, only pushing material ahead of it along fieldlines, as can be
seen in the left column of Fig.~\ref{fig3}. A plane at $z=0$ rather
than at $y=0$ reveals the same structure for all timepoints. This
structure is a result of the weak wind in combination with the
comparatively high-density filamentary cloud structure, collimating
the weak stellar wind into a bipolar outflow away from the plane of
the cloud at $y=0$. The filamentary molecular cloud structure is
relatively unaffected across the corrugated sheet of the cloud, with
only a 5\,pc diameter hole punched through the 100\,pc-diameter sheet
at the location of the star, as can be seen in the middle column of
Fig.~\ref{fig3}. Examining the movement of the stellar wind material
as shown in the right column of Fig.~\ref{fig3}, we see that very
little wind material gets more than 25\,pc away from the star.

During the red super giant (RSG) phase of evolution, from
t$_{wind}$=11.2\,Myr until its SN explosion, the slow dense wind
deposits considerable amounts of material into the cavity formed by
the earlier wind, as can be seen in the change of density around the
star between rows 3 and 4 of Fig.~\ref{fig3}. The RSG wind has a lower
energy injection rate and so the main sequence wind-blown-bubble
collapses somewhat.  There is no Wolf-Rayet (WR) stage during the
stellar evolution of the 15\,M$_\odot$ star \citep[cf.][]{georgy15}.

The accumulation of RSG wind material around the location of the star
affects the early evolution of the following SN, as shown in 
Figs.~\ref{fig4} and \ref{figA2}. However, since the
total mass-lost during the RSG stage ($\sim 1.5$\,M$_\odot$) is
much less than the mass of the SN ejecta, the RSG wind can only be a
small perturbation to the initial development of the SN remnant.  By
11.8 kyrs, as shown in Fig.~\ref{fig4}(b), the forward shock of the SN
is clear. The expansion is only being hindered by the comparatively
high-density corrugated sheet of filamentary molecular
cloud. Elsewhere, the forward shock is expanding most quickly into the
remains of the wind cavity, sweeping up tenuous wind material. Given
the nature of the explosive SN event, the pressure-driven forward
shock is also expanding into the lower density surroundings of remnant
diffuse cloud, sweeping up material from the cloud ahead of the shock.
The high-density filamentary structure is beginning to be ablated by
the SN remnant, mixing molecular, wind and SN material in flows away
from the stellar location.  In Fig.~\ref{fig4}(c) the more rapid
expansion through the wind cavity is clearer, as the thin shells
sweeping up the tenuous wind material are furtherest from the stellar
location by this time, 50 kyrs after the SN event. The sweeping up of
the magnetic field lines as the shell expands is also clear.  Panels
(d), (e) and (f) of Fig.~\ref{fig4} show the continued expansion of
the SN remnant with ablation of the molecular cloud until it reaches
the edge of the computational domain.  The progression of shocks
internal to the SN remnant are particularly clear comparing panels (e)
and (f), carrying material away from the ablating molecular cloud and
introducing internal density structure to the SN remnant.

\begin{figure*}
\centering
\includegraphics[width=175mm]{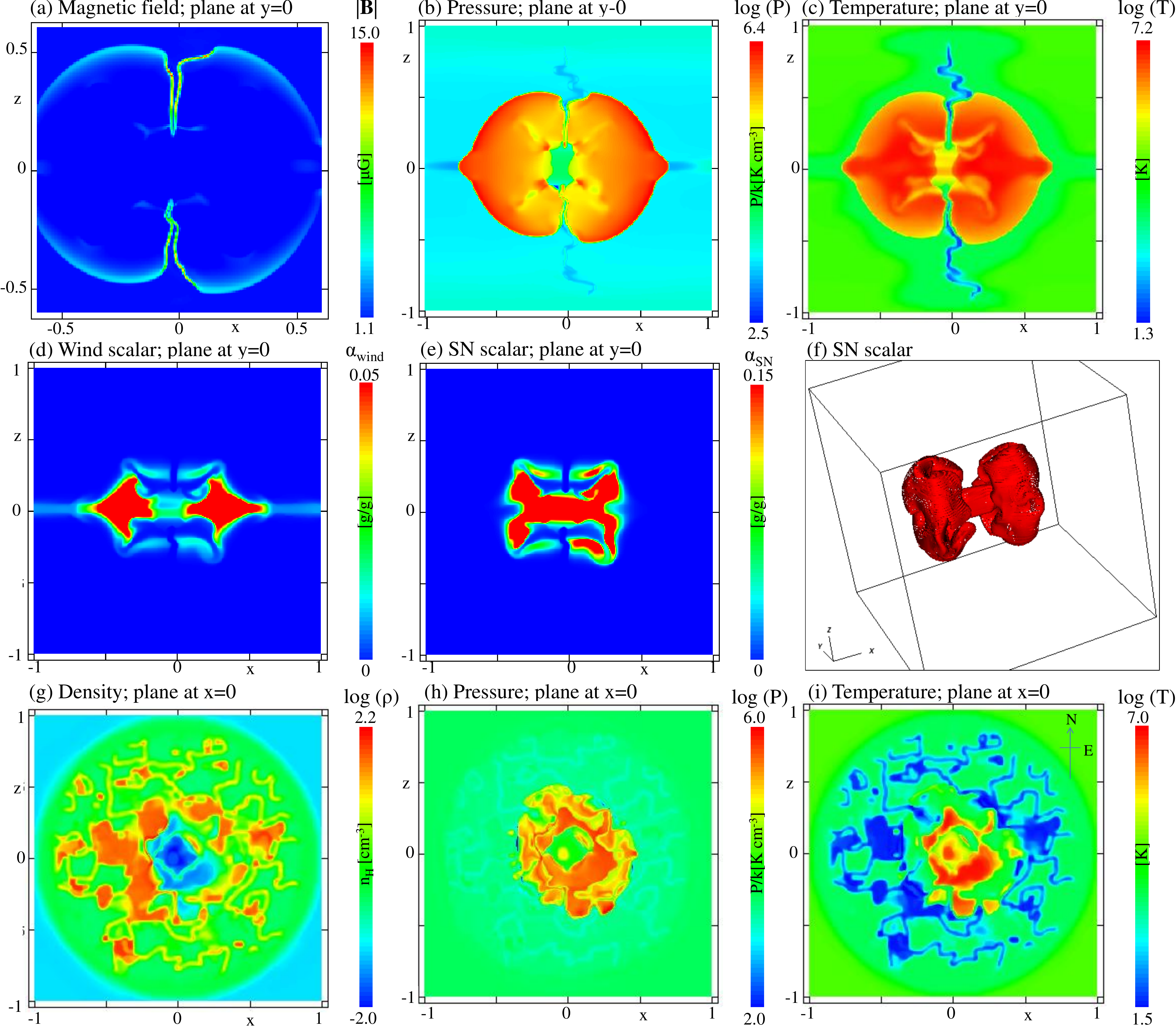}
\caption{Details of the SN-wind-cloud interaction\,50.1 kyrs into the SN phase of a 15 
M$_\odot$ star. Length is scaled in units of 50\,pc.} 
\label{fig5}
\end{figure*}

In Fig.~\ref{fig5} we show more detail of the SN remnant at one point
in time, 50.1\,kyrs after the SN explosion, as previously shown in the
third row of Fig.~\ref{fig4}. In Fig.~\ref{fig5}(a) we show the
magnetic field strength, detailing the enhancements in the shell and
at the edges of the filamentary structure alluded to in the previous
paragraph. Magnetic field strengths in the swept up shell have
increased from the initial uniform 1.15\,$\mu$G by up to a factor of 4
around the shell away from the filamentary cloud. At the edges of the
corrugated sheet of the filamentary molecular cloud structure,
magnetic field strengths have increased by a factor of 10 or more, up
to 16\,$\mu$G at most. This strengthening is not confined to the
ablating edge of the molecular cloud closest to the star location, but
extends and strengthens along the filament towards the location of the
SN forward shock. The correlation with adiabatic
compression of material would suggest this amplification is due to 
adiabatic compression. Generation of shear flows by the passage of 
the SN shock may also have contributed and led to the highest fields
observed in specific locations. There is no evidence in this
simulation of small-scale dynamo effects.

It is thus apparent that SN events in molecular clouds
can increase magnetic field strengths, through shock-driven
compression of the molecular cloud itself, in line with observed
increased magnetic field strength in molecular clouds. Examining the
evolution of the field before and after this time, it is apparent that
the magnetic field strengthens in the filamentary molecular cloud
from 2-3\,$\mu$G a short time after the SN, to the temporary peak 
of 16\,$\mu$G, reducing to at most 11\,$\mu$G 191.5\,kyrs after the SN
(Fig.~\ref{fig4}(d)) and then 7.4\,$\mu$G 309.5\,kyrs after the SN
(Fig.~\ref{fig4}(e)). Localised regions of increased field strength
then persist in the filamentary molecular cloud, decreasing from
7-8\,$\mu$G at this time until long after the SN event and wider
destruction of the cloud occurs, discussed later.  Temporary localised
increases in magnetic field strength around the edge of the expanding
SN remnant are briefly apparent, reaching up to 8\,$\mu$G. It is
possible to conclude that SN events in molecular clouds can
temporarily increase magnetic field strengths by a factor of 10 over
timescales of 50 kyrs after the SN and create persisting magnetic
fields four times stronger than the ambient field strength
for hundreds of thousands of years after the SN.

\begin{figure*}
\centering
\includegraphics[width=175mm]{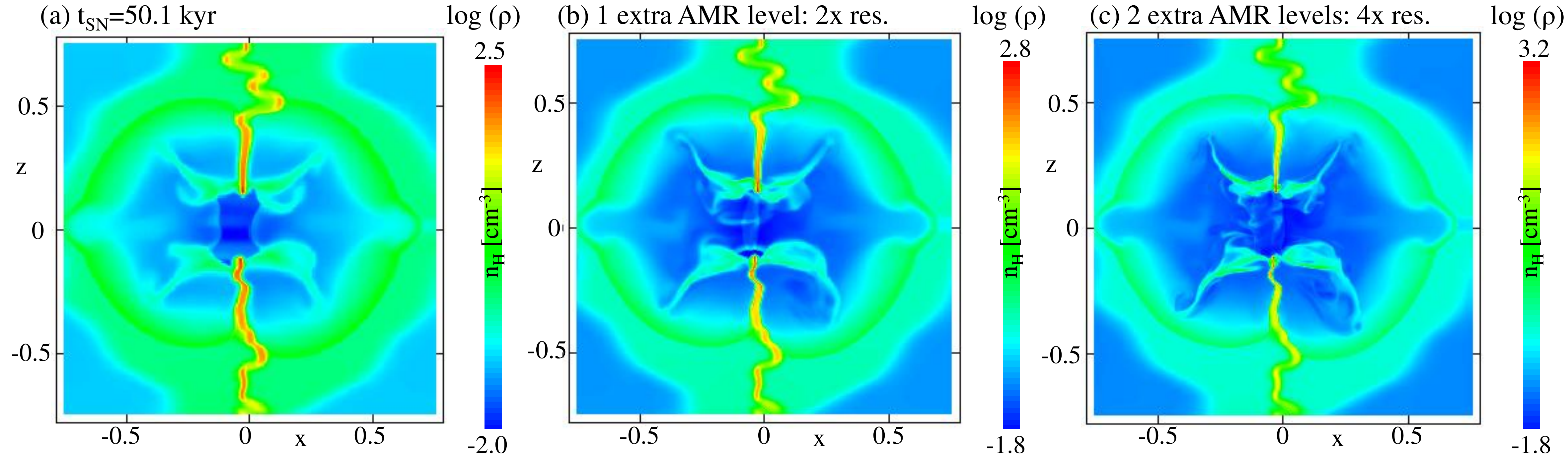}
\caption{Results of a resolution test simulating the SN phase of a 15 M$_\odot$ 
star. (a) 8 levels of AMR, as presented previously; (b) one extra level of AMR
doubling the resolution; (c) two extra levels of AMR, doubling the resolution again
from (b). Length is scaled in units of 50\,pc.} 
\label{fig6}
\end{figure*}

\begin{figure*}
\centering
\includegraphics[width=175mm]{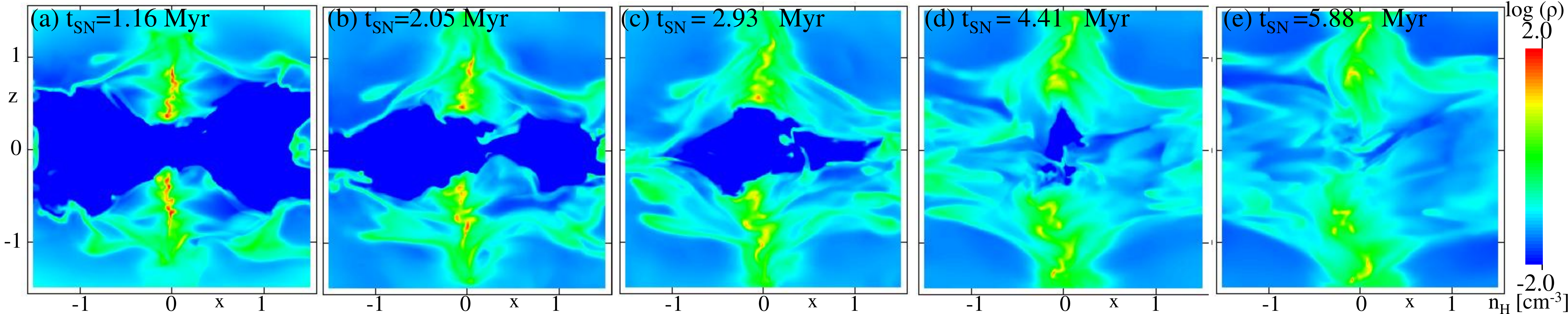}
\caption{Late SN-wind-cloud interaction during the SN phase of a 15 M$_\odot$ star. 
Shown is the logarithm of mass density on the plane at y=0 at various times through 
the SNR evolution. The SN forward shock has evolved off the grid and hence boundary
effects are now present. Ideally a larger grid is required, but the figure is shown 
to demonstrate the possible late-stage evolution of the SN-affected filamentary cloud. 
Length is scaled in units of 50\,pc.} 
\label{fig7}
\end{figure*}

In the rest of Fig.~\ref{fig5} we show the further details of the
expanding SN in order to highlight various properties of the remnant
at this time. Figs. \ref{fig5}(b) and \ref{fig5}(c) show the pressure
and temperature in the SN remnant, clearly illustrating the
pressure-driven nature of the SN at this time and the high
temperatures (up to 10$^7$\,K) behind the forward shock. The low
temperatures in the filamentary molecular cloud are also clear in Fig
\ref{fig5}(c), persisting through the passage of the SN shock. The
distribution of stellar wind material at this time is shown in Fig
\ref{fig5}(d) - note the scaling from 0 to 0.05 to elucidate the
extended distribution of wind material. The forward shock of the SN
has passed over nearly all the wind material, raising its temperature
to 10$^7$\,K, but densities are low, hence any emission from this
material will be minimal. The distribution of material ejected from
the star in the SN event is shown in Fig.~\ref{fig5}(e), again scaled
to elucidate it's furthest extent. The distribution at this time has
been shaped by the escape of ejected material from the star's location
along the wind cavity. The motion of SN ejecta directly along the $z=0$
axis has been impeded by the wind material, with which it now mixes.
Ablation of the molecular material from the
inside edges of the hole in the molecular cloud has led to the `X'
shape (missing the central section) apparent in density, pressure and
temperature plots on the $y=0$ plane and made conspicuous by the
absence of wind and SN material in the scalar plots. It should be
noted that in the arms of the `X' is hot ($>$10$^5$\,K),
comparatively high density ($>$10 cm$^{-3}$) material that originated
in the molecular cloud and that is likely to be emitting relatively
strongly in a SN remnant such as this.

The volume rendition in Fig.~\ref{fig5}(f) shows the full extent of
the ejected SN material - reminiscent of a mushroom cloud reflected
about the plane of the molecular cloud ($x\sim0$), with `dimples' at
the top of the cloud showing the impeding effect of the swept-up
stellar wind material. Shown in panels (g), (h) and (i) of
Fig.~\ref{fig5} are the density, pressure and temperature on the $x=0$
plane, to show the extent to which the expanding SN remnant has
affected the corrugated sheet of the molecular cloud.  Compared to
plots of density on the same plane in Fig.~\ref{fig3}, there is little
to no difference outside the SNR, where the SN has compressed and
heated some of the molecular material. Placing a compass on the figure
(as shown in Fig.~\ref{fig5}(i)), in the regions to the south-east the
SN has expanded the furthest and raised the pressure and temperature
of material swept up by the forward shock. This is not surprising, as
consideration of the same region in Fig.~\ref{fig3} shows the lowest
density structure in this direction and hence naively the least
resistance to an expanding SN shock.

In order to reassure ourselves that we are resolving the cooling
lengths in this simulation, we have checked that this is the case and
whilst the resolution is close to the cooling length in the expanding
SN shock, the cell size is always less than the cooling length.  To
reassure ourselves further, we have also performed extra, high
resolution simulations of the very early SN phase, starting from the point at which the
SN ejecta and energy have just been injected. We show the results of
this resolution test in Fig.~\ref{fig6}. In panel (a) we show the
distribution of density on a logarithmic scale as previously shown in
Fig.~\ref{fig3}, 50.1\,kyrs after the SN explosion.  We plot the
density on the same plane at the same time for a simulation with 1
extra level of AMR (doubling the resolution) in panel (b) and a
further extra level of AMR, equalling 10 in all, in panel (c),
doubling the resolution again. Whilst the qualitative details of the
internal shock structures vary within the SN remnant, it's clear that
the expanding forward shock is in the same place, as is the ablating
edge of the molecular cloud, and trails of ablated material away from
this edge. The internal structure may be more sharply 
resolved in both higher resolution cases, but this is not particularly 
important to the overall evolution of the cloud and the cost of such 
simulations prohibits execution at these higher resolutions. Each
further level of AMR introduces a computational-cost-multiplier of between
3 and 4 to the base resolution cost of 60,000 CPUhours per feedback case.
It is notable though that the extra resolution allows for
the refined capture of the compression of the corrugated molecular sheet
behind the ablating edges, reaching densities up to 1500\,cm$^{-3}$ in
the highest resolution simulation. Such densities might be high enough to  
trigger star formation at these locations.

In Fig.~\ref{fig7} we show the late time evolution of this simulation.
Note that the following results are indicative of the resulting
evolution, and should not be taken too literally, since the forward
shock has now left the computational domain and boundary effects have
come into play. Nevertheless, they can be used as a guide to the sort
of behaviour one might expect to see. We are particularly interested
in the fate of the corrugated molecular sheet. At 1.164\,Myr after the
SN (panel (a)), the filamentary molecular cloud is still reasonably
recognizable in the simulation, albeit being driven away from the
location of the SN, overcoming the self-gravity of the cloud. 
However, after 3\,Myrs (panel (c)), the molecular
cloud is clearly dispersing, with little high density structure (on
this plane) compared to earlier times, and by 6\,Myrs, the cloud has
been dispersed. Self-gravity in the dispersed clumps
is now likely to dominate their evolution, possibly leading to further
star formation. It is highly likely that the disruption of the
molecular cloud is caused by the SN event, but to be sure future
simulations with a larger computational volume are necessary. 
Further questions over the fate of the remaining molecular 
material and whether the SN triggers any further formation of cold
material by disrupting the thermal stability of the cloud material are
addressed in Section \ref{analysis}.

\subsection{Scenario 2 - a 40\,M$_{\odot}$ star}\label{res-case2}

\begin{figure*}
\centering
\includegraphics[width=155mm]{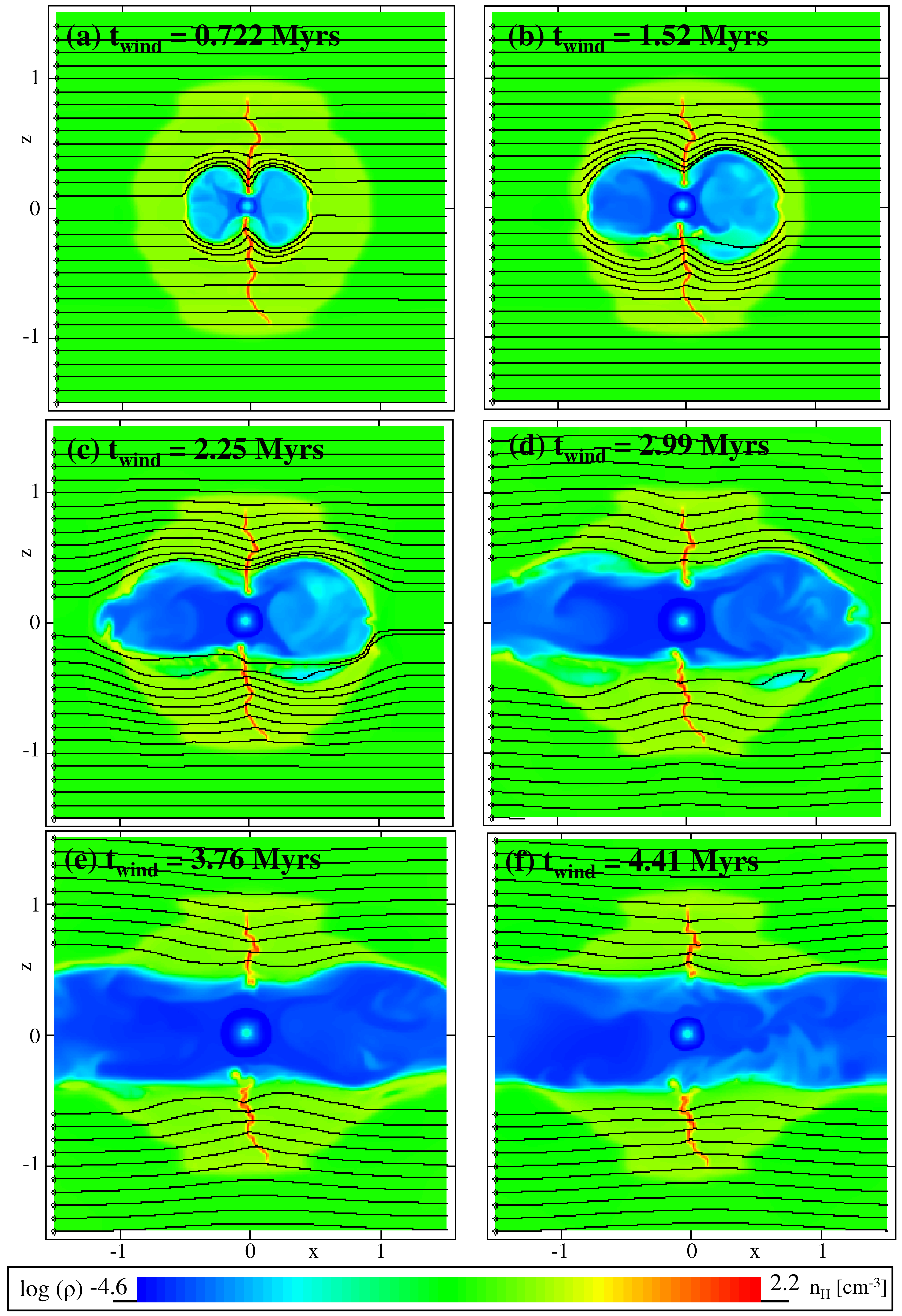}
\caption{Intra-molecular cloud evolution of the wind-blown-bubble from a 40 M$_\odot$ star. Shown 
is the logarithm of mass density on a plane at y=0. Length is scaled in units of 50\,pc.
For line profiles on these planes, see Figure \ref{figA3}.}
\label{fig8}
\end{figure*}

In Fig.~\ref{fig8} (see also Fig.~\ref{figA3}) we show the logarithm of density on a plane at
$y=0$ including the location of the 40\,M$_{\odot}$ star. At only
0.722\,Myrs into the main sequence evolution of the star, the impact
on the molecular cloud is significant and clearly different from the
15\,M$_{\odot}$ star case, as shown in Fig.~\ref{fig8}(a). The stellar
wind is forming a bipolar cavity around the location of the star, with
a narrow-waist coincident with the corrugated sheet structure of the
molecular cloud. This time, the wind is powerful enough to sweep up
diffuse cloud material ahead of it and affect the magnetic field,
trebling the magnetic field strength in the swept up diffuse cloud
material. Again, indications are that this is due to
adiabatic compression. No magnetic field is injected with the stellar wind in these
simulations but in general the stellar magnetic field is not expected
to be strong enough to significantly affect the dynamics.

The bipolar cavity grows as the wind establishes a distinct reverse
shock by 1.52\,Myrs, shown in Fig.~\ref{fig8}(b). By 2.25\,Myrs, the
bipolar wind bubble has expanded to the edge of the diffuse cloud and
is 100\,pc across.  Magnetic field strengths have dropped to
2.5\,$\mu$G from the earlier peak. Once the stellar wind has broken
out of the diffuse cloud, it rapidly accelerates away into the
surrounding tenuous medium. This results in a cylindrical cavity
through the cloud by 3.76\,Myrs, whose diameter has only increased
slightly by 4.41\,Myrs when the star evolves off the main sequence. By
4.41\,Myrs, the corrugated sheet structure of the molecular cloud has
been driven back to the edge of the cylindrical cavity, approximately
25\,pc from the location of the star. As such, the star now exists
inside a large cylindrical tunnel-like cavity punching through the
centre of the molecular cloud. This cavity is filled with hot, tenuous
stellar wind material, moving away from the star at speeds over
1000\,km\,s$^{-1}$. Much of the injected wind material flows out of
the cavity at the edge of the grid, taking $\sim 10^{5}$ years to do
so. The orientation of this channel is stable throughout the latter
stages of stellar evolution.

\begin{figure*}
\centering
\includegraphics[width=175mm]{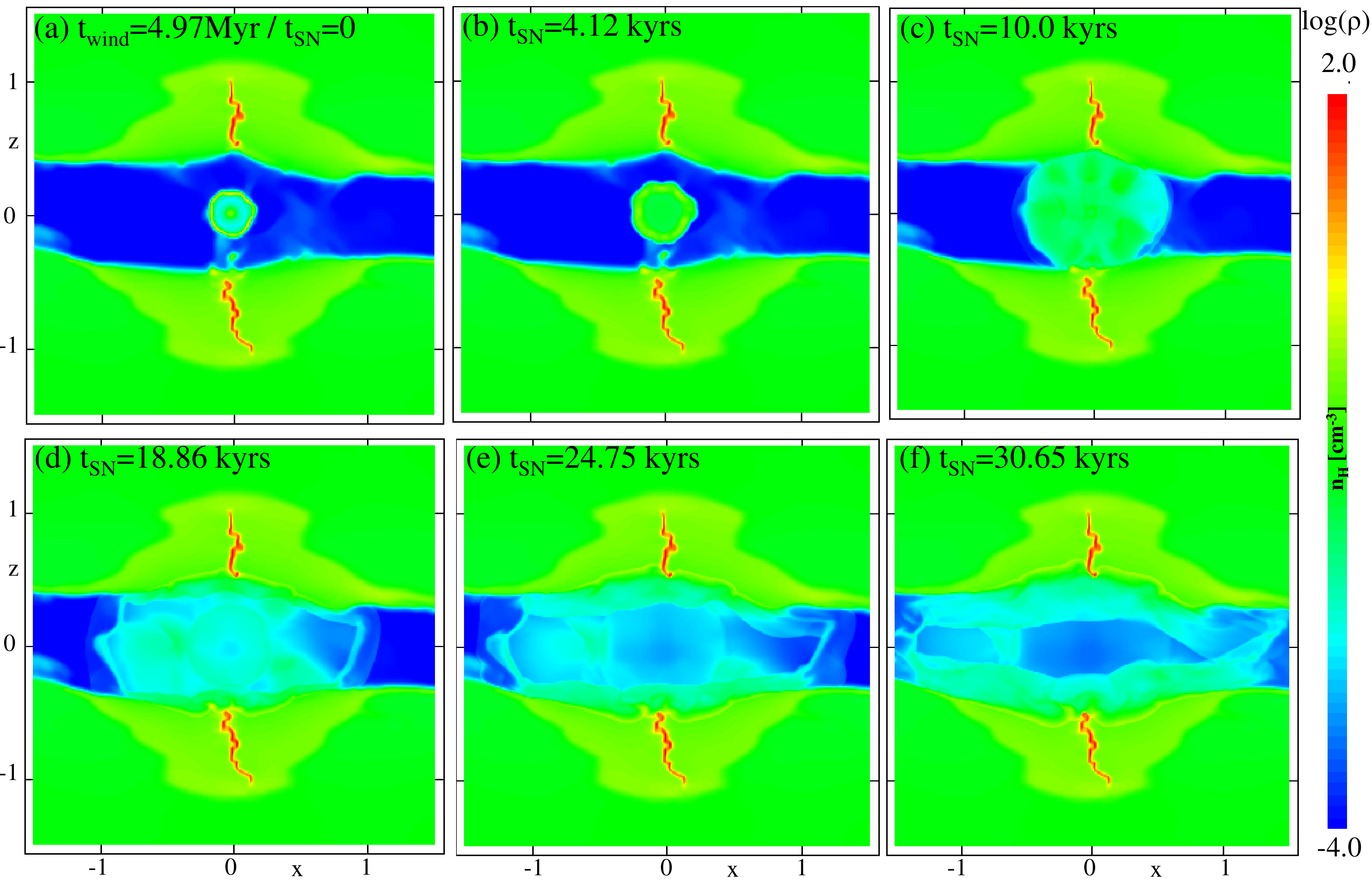}
\caption{Intra-wind cavity evolution of the SN phase of a 40 M$_\odot$ star. Shown is 
the logarithm of mass density on a plane at y=0. Length is scaled in units of 50\,pc.
For line profiles on these planes, see Figure \ref{figA4}.} 
\label{fig9}
\end{figure*}

Next, the star enters the LBV phase. The wind mass-loss rate increases
by two orders of magnitude to $\sim10^{-4}$\,M$_\odot$\,yr$^{-1}$,
and the terminal wind speed reduces to
$\sim100$\,km\,s$^{-1}$. This slow dense wind sweeps up a new shell
into the previous wind phase and forms a high density environment
around the location of the star, which eventually contains
$\sim20$\,M$_\odot$ of LBV wind material.  This phase lasts
approximately 200\,kyrs and is followed by the WR phase of stellar
evolution, where a variable, faster, less dense wind sweeps up the LBV
wind over the course of the final 400 kyrs of the star's life. The
resulting environment into which the SN mass and energy are injected
is shown in Fig.~\ref{fig9}(a). A 25\,pc diameter near-spherical high
density structure is located in the centre of the stellar wind cavity,
isolated from the molecular cloud by the 25\,pc radial extent of the
main sequence wind tunnel-like cavity.

In the rest of Fig.~\ref{fig9} (see also Fig.~\ref{figA4}), we show the evolution of the
SNR. After 4\,kyrs the SNR shock has propagated through the LBV/WR
wind-blown-bubble material and is beginning to interact with the
tenuous wind material from the MS phase of evolution. The expanding
remnant remains almost spherical until it reaches the high density
``wall'' of the tunnel-like MS wind cavity, encountering then the high
density filamentary molecular cloud and the diffuse cloud surrounding
the location of the former star. In contrast, the SNR expands almost
unhindered along the MS wind cavity towards the edge of the
domain. After only 19\,kyrs, the SN shell has reached 50\,pc along the
cavity from the location of the SN (travelling at an average speed of
2600\,km\,s$^{-1}$.  Elsewhere, the progress of the SN forward shock
has been dramatically slowed by the denser diffuse cloud and molecular
sheet. Reverse shocks are also reflecting back from the cavity wall
towards the central axis of the cavity, progressing rapidly throughout
the cavity as shown in panels (d), (e) and (f) of Fig.
\ref{fig9}. After 30.65\,kyrs, the SN shell has reached the edge of
the computational domain, 75\,pc from the SN location, along the wind
cavity. As a comparison, the SN in the case of the 15\,M$_{\odot}$
star took 50\,kyrs to reach the edge of the diffuse cloud and a
further 400\,kyrs to expand to the edge of the same computational
domain. The introduction of the strong stellar wind from the
40\,M$_{\odot}$ star has wide-reaching consequences; the majority of
the SN energy and material is able to leave the molecular cloud
relatively unhindered, due to the cavity formed by the stellar wind.

\begin{figure*}
\centering
\includegraphics[width=175mm]{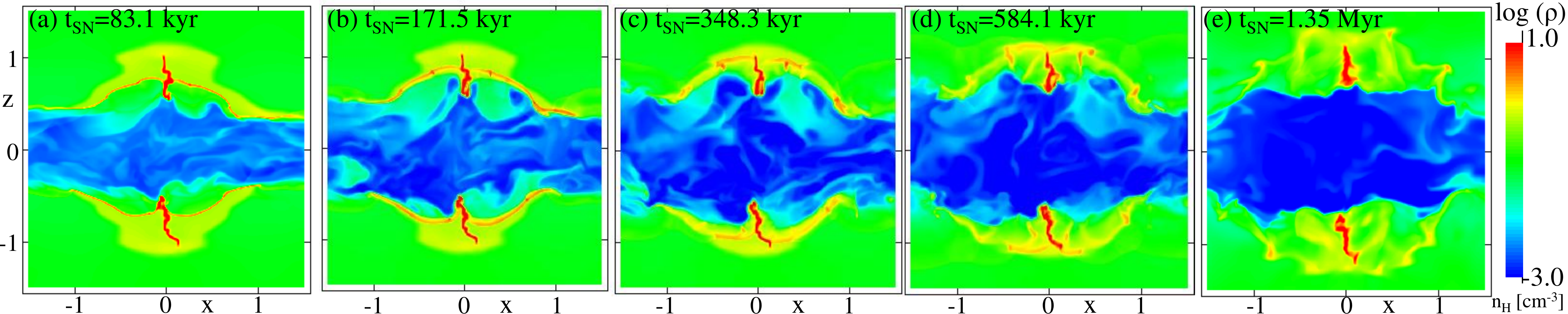}
\caption{SN-cloud interaction during the post-SN-cavity-interaction phase of a SN from a
40 M$_\odot$ star. Shown is the logarithm of mass density on a plane at y=0. Length is 
scaled in units of 50\,pc. For line profiles on these planes, see Figure \ref{figA5}.} 
\label{fig10}
\end{figure*}

During this short intra-cavity stage, the SN remnant also begins to
expand into the filamentary molecular cloud and its parent diffuse
cloud. It takes considerably longer to pass through these regions -
1.35\,Myrs. We show snapshots of this period in Fig.~\ref{fig10}
(see also Fig.~\ref{figA5}). 83\,kyrs after the SN explosion, as shown in
Fig.~\ref{fig10}(a), the SN shock is sweeping up diffuse cloud
material ahead of it, leaving material behind the shock to flow into
the cavity. The flow along the cavity and off the grid is primarily
driven by the earlier expansion of the remnant out of the cavity, but
also by reverse shocks off the cavity walls. This effect is further
pronounced in Figs. \ref{fig10}(b) and \ref{fig10}(c), 171.5 and
348.3\,kyrs after the SN explosion. The slowing down of the forward
shock as it sweeps up the diffuse cloud between these times is also
apparent.  By 584\,kyrs after the explosion (Fig.~\ref{fig10}(d)), the
SN remnant has expanded and cooled far enough to reach the radiative
phase of evolution. The shell becomes subject to Rayleigh-Taylor instabilities
and breaks up into individual clumps, as can be seen most clearly
in Fig.~\ref{fig10}(e), 1.35\,Myrs after the initial SN event.
Interestingly, the filamentary molecular cloud survives this phase of
its evolution remarkably intact, apart from the large hole in the
centre of the corrugated sheet that was originally caused by the
stellar wind. This hole has expanded in diameter by 20\% during its
interaction with the SN. The corrugated sheet has also expanded
outwards, driven by the SN, but has retained its form and high
density.  Magnetic fields in and around the filament and at the edges
of the SN remnant are temporarily increased by factors of 3 or 4, but
no more, and by 1.35\,Myrs have returned to around 2\,$\mu$G, having
started out in the diffuse cloud at 1.15\,$\mu$G.

The accuracy of this phase of the simulation can be questioned, due to
the SN shock having expanded along the cavity off the grid. Progress
of the SN shock into the diffuse cloud and filamentary sheet does not
appear to be affected by boundary effects that may have affected the
evolution of the intra-cavity material, and for this reason we can be
reasonably confident in the description of the evolution until 1.35
Myrs post explosion.  After 1.35\,Myrs, the SN evolution follows a
path remarkably similar to the 15\,M$_{\odot}$ star case, refilling
the domain and eventually disrupting the host molecular cloud on the
same 6\,Myr timescale. It is likely that in this case, physical causes
include the interactions of the shocks that bounced off the inside of
the cavity and back towards the location of the star. These then
progress across the whole cavity and generate further shock echoes,
which all drive weaker yet destructive shocks into the molecular
cloud. Gravity also begins to play a role at this time, but
because boundary effects will progressively affect the
simulation past this point we do not show any further details, which
will require additional larger domain simulations.

\section{Analysis}\label{analysis}

\subsection{Energy}
\label{energy}

\begin{figure*}
\centering
\includegraphics[width=165mm]{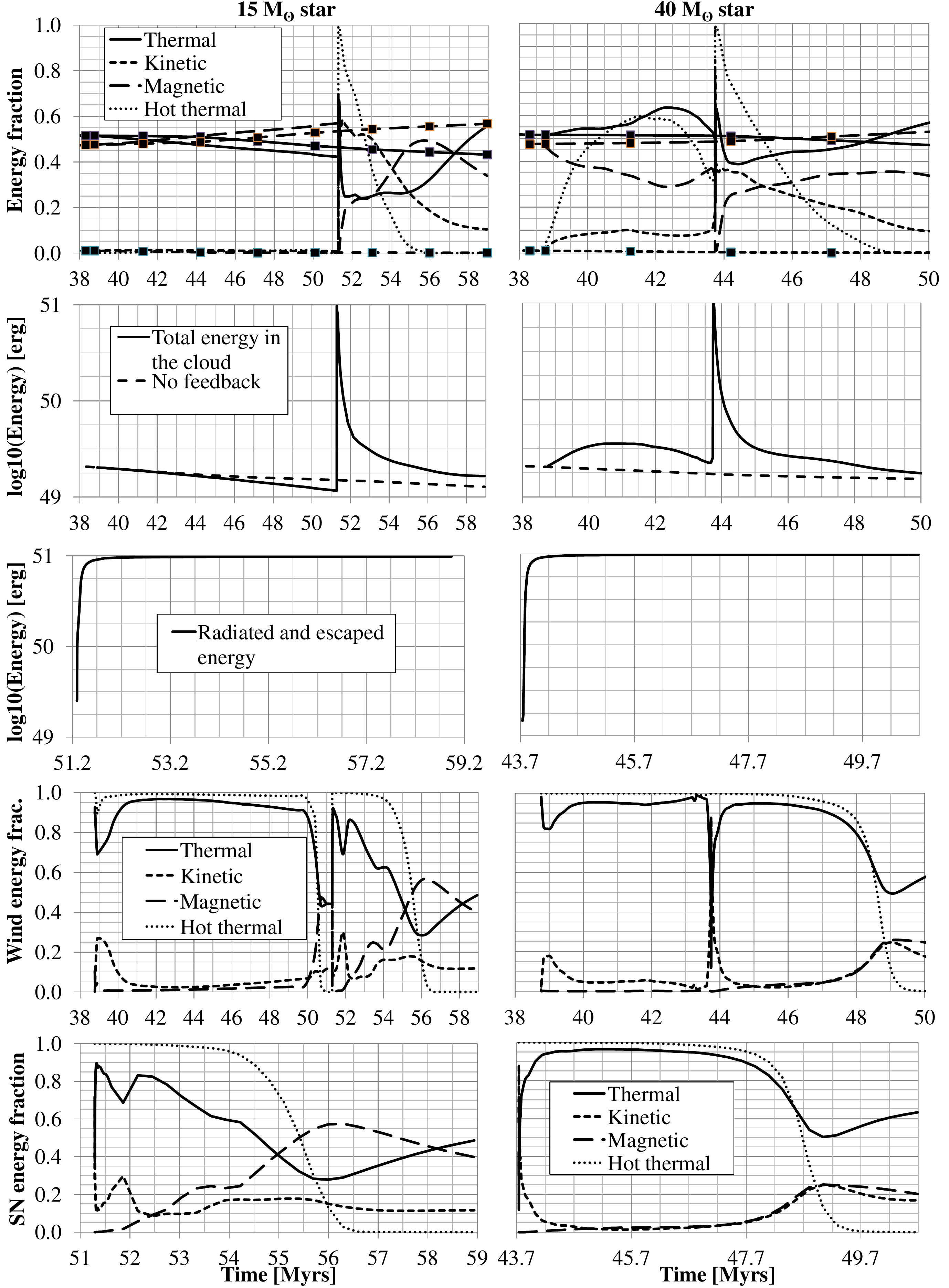}
\caption{Time profiles of energy fractions, totals and escaped amounts
  of energy in the cloud and separately for wind energy and SN
  energy. The thermal, kinetic and magnetic energy fractions together
  sum to 1.0 i.e. we are ignoring the change in gravitational energy. The 
  `hot thermal' profile indicates the fraction of the thermal energy which 
  is hot (above $10^{4}$\,K), and
  so also has a range between 0.0 and 1.0. In the first row, lines
  with markers indicate the energy behaviour in the cloud with no
  stellar feedback. In the third row, the ``radiated and escaped energy"
  is the total cloud energy subtracted from the injected energy of the SN, 
  10$^{51}$ erg.}
\label{fig11}
\end{figure*}

Across the first row of Fig.~\ref{fig11} we show how the distribution of energy between 
kinetic, magnetic, thermal and hot thermal (i.e. material above 10,000K) varies 
over time following the introduction of stellar feedback, with reference to a 
simulation of the cloud evolution without feedback (indicated by lines
with markers). The thermal, kinetic and magnetic energy fractions
together sum to 1.0 i.e. we are ignoring gravitational energy.
 
In the 15\,M$_{\odot}$ star case, the fractions do not change considerably during 
the wind phase, as compared to the reference case. This was expected from the
minimal dynamic and structural impact the star has on the cloud. It is also supported 
by the fact that the majority of the energy in the wind is thermal, as shown in the 
fourth row of Fig.~\ref{fig11}, highlighting the minimal kinetic energy introduced by the 
relatively weak stellar wind. In fact the star introduces so little energy, that the overall
energy in the cloud reduces during the wind phase, as compared to the `no feedback'
reference case. This likely arises due to a combination of: i) some
small fraction of material leaving the cloud; ii) relatively efficient
radiation of the energy injected by the stellar wind, perhaps due to
weakly compressing and heating neighbouring cloud material.

The introduction of the thermalised SN kinetic energy is most obvious in the total energy
plot, as the discontinuity at 51.3\,Myrs. Since the SN energy is
injected thermally, both the thermal energy and the `hot thermal'
phase spike at this time. Over time, the thermal fraction drops 
off to 25\%, as some of the SN thermal energy transforms into kinetic
energy, and as the SNR cools. For the first $10^{5}$ yrs, the SNR is
expanding into diffuse cloud with $n_{\rm H}\sim 1\,{\rm cm\,s^{-1}}$
and into the cold corrugated sheet with $n_{\rm H}\sim 100\,{\rm
  cm\,s^{-1}}$. The timescale for the transition of a
SNR into its radiative phase was determined by \cite{blondin98} to be
$\approx 1.5 t_{\rm tr}$, where $t_{\rm tr}$ is given by their Eq.~3. For
these densities, we expect the remnant to become radiative after
$\approx 5\times10^{4}$\,yr expanding into the diffuse cloud and 
$\approx 4\times10^{3}$\,yr expanding into the cold corruagted sheet.
This is in keeping with our simulation results: in the ``radiated and escaped 
energy" plot, 80\% of the SN energy radiates away prior to the SNR
starting to leave the 
grid ($\approx 0.3$\,Myrs after the explosion, or 51.6\,Myrs since the
start of the simulation).

Once the SN remnant
has left the grid, it is no surprise that the kinetic energy fraction falls against the rising
thermal and magnetic energy fractions (although with no injection of magnetic energy, 
in reality the total magnetic energy is either maintained or decreases over time as
material leaves the grid). The variations of energy fraction during the SN phase are
most clearly shown in the plot of the SN energy fraction on the fifth
row. We remind the reader that all plots should be interpreted with caution at times after the
SNR first reaches the boundaries of the numerical domain.

For the 40\,M$_{\odot}$ star case, the stellar wind has a considerably larger effect on the
energy fractions. The star introduces a large amount of hot thermal and kinetic energy,
raising both fractions considerably above the reference case. The magnetic energy fraction
goes down, but this reflects an increase of total energy in the simulation during the wind
phase (as shown in the second row) rather than a reduction of magnetic energy. Given
that the wind blows a hot bubble, it's likely that radiative losses are responsible for
the drop in total energy approaching the end of the star's life, before the introduction
of 10$^{51}$\,erg in the SN event. In this case, the SN would appear to have less of an
effect on the energy fractions than the 15\,M$_{\odot}$ star - no surprise given the ease
with which SN ejecta leave the cloud along the tunnel. Closer inspection has shown that
there is a spike in kinetic energy up to 0.8 for 30,000 years post-SN as the SN material
is ejected down the tunnel and out of the cloud. This fraction drops rapidly as the SN shell 
expands into the cloud, as shown in the second row and in detail in the final row of the figure,
where the initial kinetic energy peak is clearer. 5\,Myrs post-SN, the energy balances
change considerably from their previous post-SN behaviour. It is most likely that this is 
influenced by boundary effects after the forward shock of the SNR has 
crossed the grid boundary. As such, we again reiterate that results late after the SN must be treated with
caution.

\subsection{Maximum density}

\begin{figure*}
\centering
\includegraphics[width=165mm]{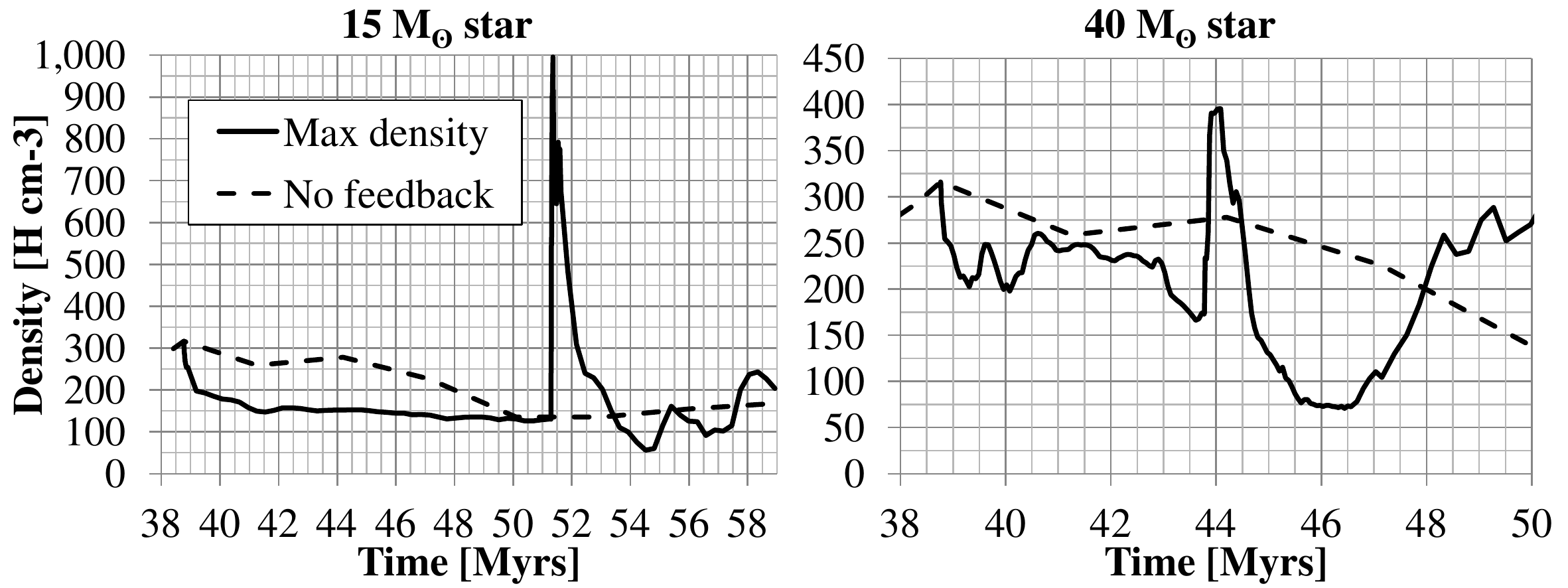}
\caption{Maximum density in the two stellar feedback simulations (solid lines) 
shown with the maximum density in a continued simulation of the same initial conditions without 
feedback (dotted line) for reference.} 
\label{fig12}
\end{figure*}

In Fig.~\ref{fig12} we show the maximum densities reached in the simulations,
compared to the maximum density at the same time in the reference case without
stellar feedback. In the 15\,M$_{\odot}$ star case, the maximum density during
the wind phase is considerably lower than the reference case, until just before the SN.
Since we chose to position the star and inject the wind at the highest density location close
to the centre of the cloud, this behaviour reflects that choice. The final wind
stages inject a considerable amount of mass at low velocity around the star's
location, raising the density back to `no feedback' levels. The SN instantly raises
the density in the injection location, but the spikes seen in the plot are not
associated with the SN injection, but instead with compression of the filamentary
cloud by the SN shock, leading to maximum densities far above the `no feedback'
case for nearly 2\,Myrs after the SN event. Later peaks in the maximum density
plot for the 15\,M$_{\odot}$ star may be associated with the formation of new
cold material after SN affected material returns to the thermally unstable
phase. These peaks again cross the threshold at which we formed our star and at
which other authors inject stars, which if incorporated would
`trigger' further star formation.

The plot of maximum density for the 40\,M$_{\odot}$ star shows a much more
variable maximum density during the wind phase. Initially this is less than the reference
case, but at times it comes close to returning to the reference case. The variations
are not due to the variations in the stellar wind - it is relatively
steady for 4\,Myrs as shown 
in Fig.~\ref{fig2} - but instead are due to the compressional effect the stellar wind 
has on the cloud, creating first the bipolar bubble (generating the spikes in
maximum density) and then blowing out the tunnel through the cloud (reducing the
maximum density once the tunnel has become established). The final stellar
wind stages and the SN event are the cause of the peak in maximum density at 
44\,Myrs. There is not enough energy to drive as high a peak as in the 15\,M$_{\odot}$ 
star case; a lot of the energy is `vented' along the tunnel and
therefore there is less compression 
of the gas. The rise in density is sustained for a period after the SN event of 
approximately 0.5\,Myrs, before maximum densities in this case drop far below
the reference case. Whilst a significant fraction of the SN ejecta may have left the cloud
along the tunnel, clearly the SNR still has a strong effect
on the parent cloud over the next 3\,Myrs, reducing densities far below the reference case
and based on our criterion, turning off any further star formation. After this time, maximum 
densities in the simulation begin to rise again. This correlates with the increase in cold 
material triggered by both the wind and the SN forcing material back into the
thermally unstable phase, discussed in more detail in the next section.

\subsection{Phases}

\begin{figure*}
\centering
\includegraphics[width=165mm]{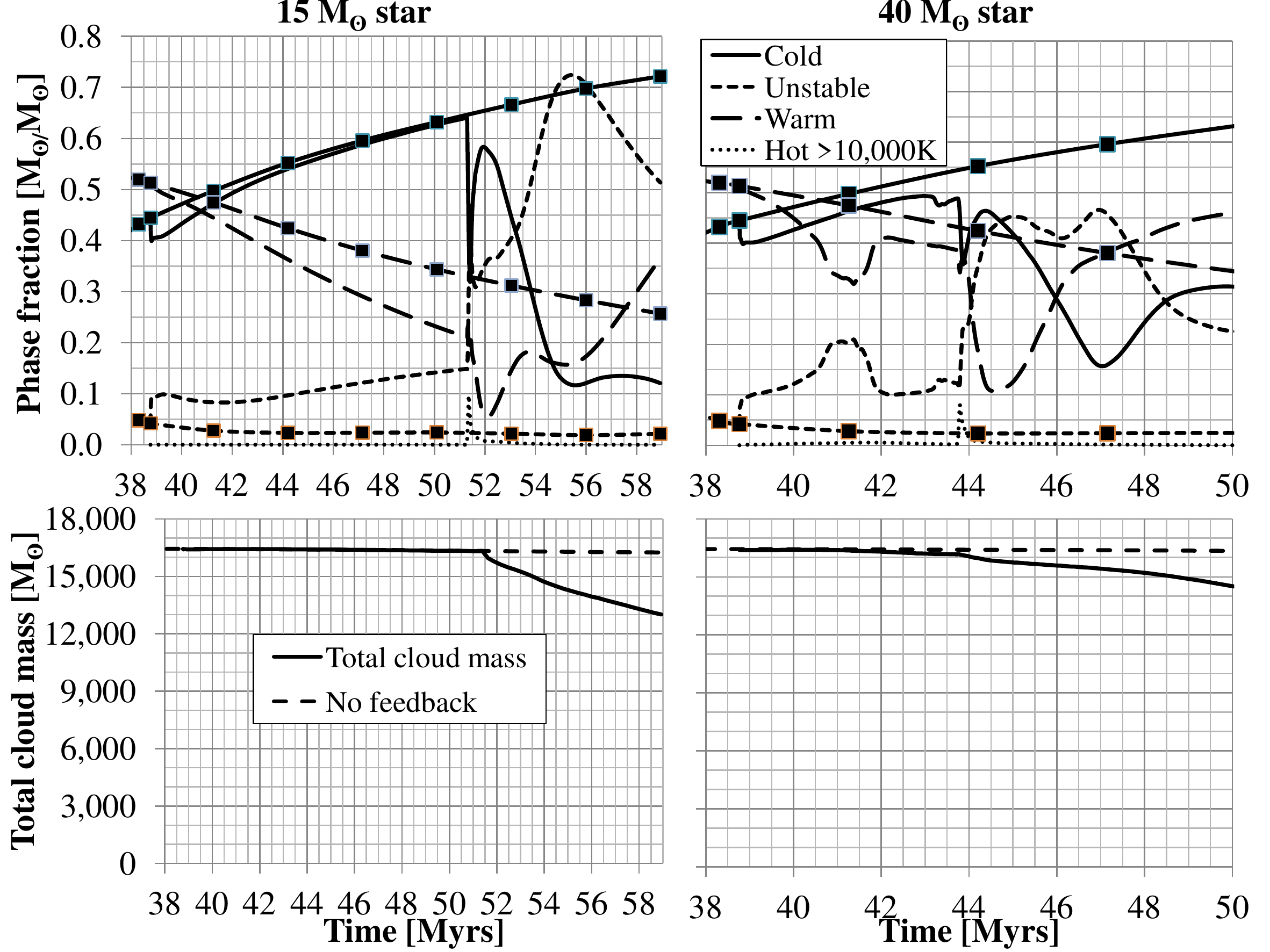}
\caption{Phase fraction and total cloud mass for the two stellar feedback simulations.
The reference case without feedback is indicated by lines with markers in the phase
fraction plots and dashed lines in the total cloud mass plots.} 
\label{fig13}
\end{figure*}

In Fig.~\ref{fig13} we show the phase fraction and total cloud mass for the 
two feedback simulations. Prior to feedback beginning, the cloud is thermally
in equilibrium with 50\% of its mass in the warm phase (with temperature between
5,000\,K and 10,000\,K) and $\approx 45$\% of its mass in the cold phase 
(with temperature below 160\,K). The remaining 5\% is in the unstable equilibrium 
part of the pressure-density phase-space between 160\,K and 5,000\,K. Once mechanical feedback starts,
the hot shocked gas moves out of thermal equilbrium and takes a
cooling time to radiate away its excess energy. The low density shocked
stellar wind gas has a comparatively long cooling timescale, while the
denser and cooler swept-up gas has a much shorter cooling
timescale. Similarly, the hot SN ejecta and the gas it sweeps up have
comparatively long cooling timescales (see also the earlier discussion
in Section~\ref{energy}).

Fig.~\ref{fig13} shows that during the wind phase of the
15\,M$_{\odot}$ star case, the fraction of cold material increases
very much like the reference case with no feedback - the cloud
continues to form. However, the introduction of stellar feedback has
caused the amount of warm thermally stable material to decrease as
compared to the reference case. The temperature of this material has
presumably been increased by the compacting and pressurising effect of the stellar
wind, thereby changing it back into the thermally unstable state
(between 160\,K and 5,000\,K). The presence
of a larger quantity of unstable material as compared to the reference
case enables the condensation of more cold material, allowing the
fraction of cold material to return to the level without feedback.

Since the 15\,M$_{\odot}$ star only introduces a few solar masses 
of material during the wind phase, the total cloud mass barely changes 
until the SN. The SN then launches hundreds and eventually thousands 
of solar masses of material out of the cloud (as seen in the bottom left 
panel of Fig.~\ref{fig13}). The SN rapidly heats nearly 10\% of the 
cloud material into the hot thermal phase (above 10,000\,K), but the 
majority of this material radiatively cools quickly and the fraction of hot 
material drops back close to zero before the SNR leaves the grid (at
51.6\,Myrs as noted in Section~\ref{energy}). The amount of cold material decreases very 
rapidly post-SN, returns to a high fraction of 0.58 after 0.5\,Myrs - the 
point at which the SN forward shock has progressed through the cloud 
- and then decreases as the remnant evolves. These changes can be 
understood as the sweeping up and ejection of diffuse material from the 
cloud by the SN, whilst the cold filamentary
material reacts much more slowly. Some of this behaviour may also be
due to the (SN) shock heating and subsequent cooling of dense
material, as found by \citet{rogers13}. 

Eventually though, the cold dense material is stripped out of the
cloud. This reduction in dense cold material is matched by an increase
in lower density unstable material, indicating the disruption of the
cloud by the SN. At 4\,Myrs post-SN, around 55\,Myrs into the
evolution of the cloud, the fraction of unstable material peaks. 

At this time only $\sim$15\% of the cold phase cloud remains. However,
85\% of the total cloud mass remains on the grid. Thus the majority of
the mass remains in the cloud, though now mostly in the thermally unstable phase.
Some cold material does survive the SN, and there is evidence for a slight increase
6\,Myrs post-SN. However, this result is at the limit of applicability
of these simulations, as numerical issues are present due to the
forward SN shock having long since passed off the grid. Confirming an
increase in cold cloud material - or in other words new cloud
formation - requires further study.

To summarize, for the 15\,M$_{\odot}$ star case, the wind phase 
has very little effect on the cold cloud material - there is little 
variation in the cold phase fraction compared with the `no feedback'
case. It is only after the SN that the cloud is strongly dissipated.

In the 40\,M$_{\odot}$ star case, the fraction of cold material is
  initially reduced by the injection of the stellar wind but then
  gradually rises until the late stages of the star's evolution, when
  the large effect of the stellar wind has led to considerable
  differences as compared to the no feedback case. During the time
  taken to establish the tunnel through the cloud, between the initial
  formation of the star at 38.8\,Myrs, and the time when the wind
  bubble reaches the grid boundaries at $\approx 41$\,Myrs, the
  structural changes driven by the stellar wind significantly affect
  the diffuse parts of the parent cloud. Once the wind tunnel through
  the cloud has been firmly established (from $\approx 42$\,Myrs to
  the SN explosion at 43.7\,Myrs), the fraction of cloud material in
  the unstable phase is roughly constant, whilst the warm and cold
  material fractions change. The fact that the rates of change of the
  fractions of warm and cold material match the `no feedback' case
  indicate that this is due to continuing cloud evolution in the parts
  of the cloud that are relatively unaffected by the stellar wind. The
  total mass of the cloud slowly reduces, especially once the tunnel
  has become established, as the flow through the tunnel entrains some
  of the cloud material from the tunnel walls and carries it off the
  simulation grid.

The behaviour at and after the SN is similar to the 15\,M$_{\odot}$
case albeit with
less extreme fractional changes, in accordance with a significant part
of the SN energy leaving the grid along the wind tunnel. After a similarly 
short period, the hot phase material has cooled and the hot fraction 
has returned to zero. After 3.5\,Myrs post-SN, the cold and warm fractions
again rise, with material moving out of the unstable phase. However,
since the grid boundaries may have affected these results this observation should again
be treated with caution and requires further investigation.

\begin{figure*}
\centering
\includegraphics[width=160mm]{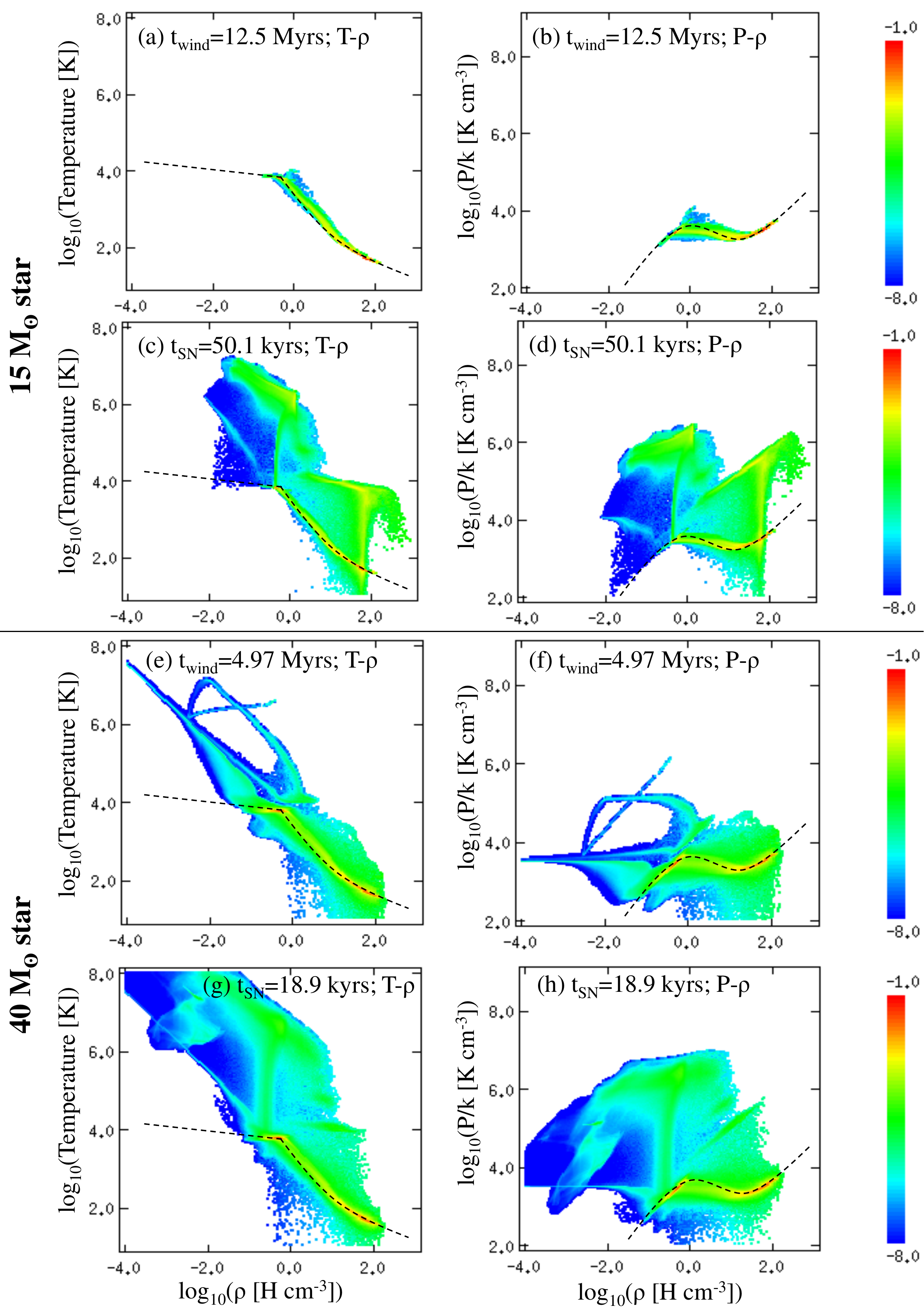}
\caption{Mass-weighted phase space probability distributions for the two simulations
presented here at different times during the stellar evolution. Over-plotted dashed
lines indicate the approximate thermal equilibrium between heating and cooling 
prescriptions used herein.} 
\label{fig14}
\end{figure*}

In Fig.~\ref{fig14} we plot the temperature-density (in the left column) and 
pressure-density (in the right column) distributions for the two cases considered
in this work using 200 bins in log density and 200 bins in log temperature/pressure.
We also over-plot the approximate thermal equilibrium curve for the heating and 
cooling prescriptions used in this work. In the first row, we show the distributions
at the end of the wind phase for the case of the 15\,M$_{\odot}$ star. 
Most of the material remains in thermal equilibrium,
tracing the equilibrium curve. Two distinct stable phases (warm and cold) are in
approximate pressure equilibrium with each other.

In the second row, we show
the distributions 50\,kyrs after the SN. Towards the upper left of each distribution
is the low density, hot phase, which consists of SN shock heated gas. It is not yet
in pressure equilibrium with the warm phase, but evolves towards this equilibrium
as the simulation progresses beyond that shown here. The distribution at low
density is reasonably wide in pressure and temperature. The reason for this is the
presence of both shock heated and adiabatically cooled gas, as noted in 
a similar analysis of such distributions influenced by feedback
simulated by \cite{walch15b}. The SN has also shock-heated an appreciable amount
of the cold dense material, spreading the distribution upwards at high density in both
plots.

In the third row of Fig.~\ref{fig14}, we show the distributions for the case of the
40\,M$_{\odot}$ star at the end of the wind phase. The distributions are considerably
different to the lower mass star case. The two distinct stable cloud phases are still present
in thermal equilibrium, but are now accompanied by the low density tunnel indicated by
the branch stretching away from the equilibrium curve horizontally to the left at the 
same P/k$\sim10^{3.5}$\,K\,cm$^{-3}$ in the pressure-density distribution. The 
"bubble within a bubble" that is the wind-blown shell around the star that formed during
the last stages of stellar evolution is responsible for the branch and arc of material out of 
equilibrium at higher pressure. Specifically, the diagonal line that spans from 
(P,$\rho$)=($10^6$,$10^{-1}$) down to connect to the tunnel branch at 
(P,$\rho$)=($10^{3.5}$,$10^{-2.75}$) is the wind injection region 
and region of undisturbed wind material up to the reverse shock. The reverse shock is 
indicated by the jump back up in pressure to the over-pressured (with respect to the 
cloud and tunnel) ``bubble within a bubble" that is itself indicated by the horizontal branch 
at P/k$\sim10^{5}$\,K\,cm$^{-3}$ that arcs back to the equilibrium curve. The greater
power of the wind has also driven a broadening of the distributions, with heated material
above the equilibrium curve radiatively cooling towards the equilibrium curve. Cooling 
by expansion has led to low density, cool gas below the equilibrium curve.

Similarly to the
15\,M$_{\odot}$ star case, the SN event has created a low-density, hot phase in the upper
left of the distribution and this hot gas component is considerably more pronounced
(the younger age of the remnant in the 40\,M$_{\odot}$ star case means that the maximum
temperature and pressure on the grid is higher). 
Unlike the previous case though, the SN has had far less effect on 
the cold material. Only late post-SN is there any sign that the SN has heated the dense
material, with less of an effect than the previous case. Eventually, the hot gas pressure
adjusts to the local pressure of the cloud, as is expected on larger scales for a given
galaxy following the effect of supernovae \citep{ostriker10}, again lending credibility to 
our simulations.

\subsection{Mixing}

\begin{figure*}
\centering
\includegraphics[width=165mm]{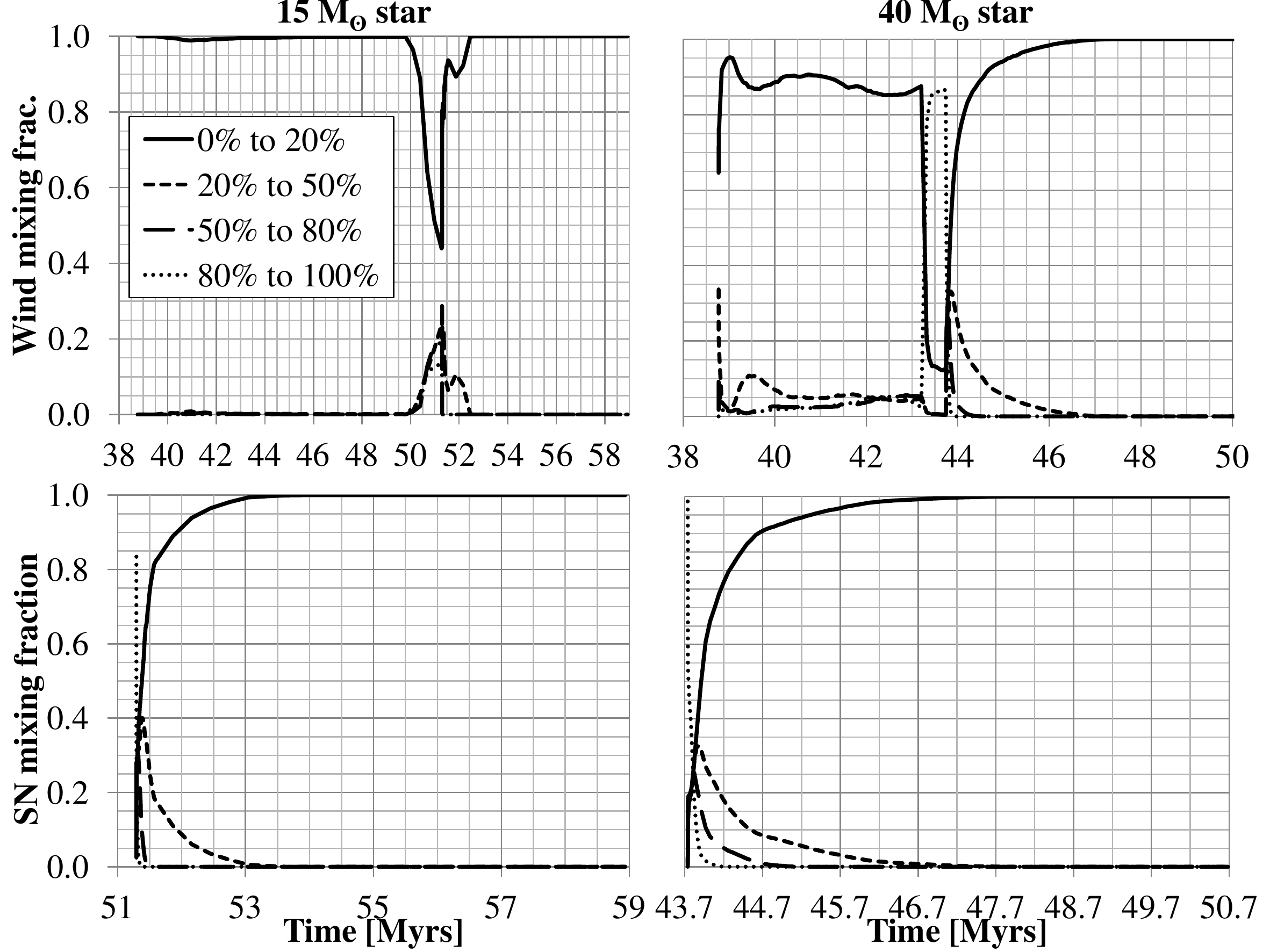}
\caption{Levels of mixing of the wind material (top) 
and SN material (bottom) with the cloud material throughout the
stellar feedback simulations. A value of 100\% indicates no mixing of the wind/SN material with the cloud
material.} 
\label{fig15}
\end{figure*}

In Fig.~\ref{fig15} we show the mixing of the stellar wind material (on the top row) 
and the SN material (on the bottom row) with cloud material. In the 15\,M$_{\odot}$ 
star case, wind material is immediately well mixed, since it is all at fractions of no 
higher than 20\% with respect to the cloud material. Whilst an equivalent amount of
mass was taken out of the injection region to make the star, some cloud material
remained in the injection region. The weakness of the wind implies that even in the very initial
stages of switching on the wind, there is no brief period when the injection region is
almost entirely wind material i.e. there is no transition observed from ``no mixing" 
(at 100\%) down to well mixed. Only during the late
phases of evolution is the slow dense wind not completely mixed
(though most of it still is). The subsequent SN then mixes well both the wind and SN material with the cloud material.
This is perhaps not surprising as neither the stellar wind, nor the SN, 
inject large amounts of material. Therefore, it can be concluded that mixing is very
efficient for both the wind and the SN associated with the 15\,M$_{\odot}$
star.

Fig.~\ref{fig15} also shows that the stellar wind of the 40\,M$_{\odot}$ star 
is not as well mixed as that of the 15\,M$_{\odot}$ star. In this case, the brief 
initial period of mixing down from almost 100\% in the injection region is indicated
by the sharp initial transience. Approximately 90\% of it is mixing down to less than 20\%
fraction of the material in a cell, but a non-zero fraction does not mix below
80\% stellar wind material. This is the material which stays in the wind tunnel
and escapes the grid along the tunnel. Again reflecting the isolating effect
of the tunnel, this time between the SN and cloud material, the SN material takes 
far longer to mix with the cloud material than in the 15\,M$_{\odot}$ star
case. However, over time, the wind and SN material post-SN gradually
mix more completely with the cloud material. Mixing for both the 15\,M$_{\odot}$
star case and the 40\,M$_{\odot}$ star case would appear to be efficient in these
simulations.

For these simulations, we have not quantitatively examined the nature
of the material leaving the grid. The wind tunnel in the
40\,M$_{\odot}$ star case does not allow for as efficient entrainment
of cloud material off the grid, but in both cases the SN drives
material out of the cloud, as investigations of the phase fractions of
material have shown in the previous section.  Further investigations
may shed more light on mass-loading and the entrainment of material
leaving the grid. We will examine this in a future work.  

\section{Comparisons to previous work and observations}\label{comparisons}

\subsection{Comparisons to previous work}

It is not immediately straight forward to compare the present work with our
previous efforts \citep{rogers13,rogers14} due to the large
differences in initial conditions. However, there are certainly some
key differences, such as the way that wind energy is transferred to
the wider surroundings (through a multitude of porous channels in
\cite{rogers13} compared to the expansion of a coherent bubble in the
present work). There are also similarities, such as the ability of SNe
to transport large amounts of energy without strongly affecting the
parent cloud (at least not quickly). In both works the key factor in
this respect is the shaping effect of the pre-SN stellar winds, which
makes the cloud highly porous \citep{rogers13}, or as in this work
opens up large-scale channels which direct the SN energy.

Other work has noted how difficult it is for SN feedback alone to
entirely disrupt more massive molecular clouds. For instance,
\cite{walch15} find that a single SN has very little impact on a
$10^{5}$\,M$_{\odot}$ cloud unless it explodes into a low density
environment. Subsequent SNe help only if the second explosion happens
during the snowplough phase of the first one. Similarly,
\cite{kortgen16} find that single SNe disperse $\sim$10\,pc-sized
regions but do not disrupt entire clouds, which instead requires
clustered, short-interval SNe to form large hot bubbles.
Our simulations focus on feedback in a cloud of lower mass than the
\cite{walch15} and \cite{kortgen16} studies, and also include the
effect of the stellar wind, so our conclusions are naturally somewhat
different. In our current simulations we also find differences in the
nature of the feedback in the 15\,M$_{\odot}$ and 40\,M$_{\odot}$ star cases,
including the initial effect of the SN on the cloud, although in both
cases the molecular sheet and the surrounding diffuse cloud are
eventually dispersed. These differences are caused by the different
structures that the SN explodes into. In the 15\,M$_{\odot}$ star case
the weak preceding wind barely creates a wind-blown cavity, and the
molecular sheet has only a small hole in its centre. The resulting SNR
sweeps up the molecular sheet but also expands into the diffuse cloud
in which the molecular sheet resides (see Fig. \ref{fig4}). The
molecular sheet is pushed back and disrupted over the following
several Myr (see Fig. \ref{fig7}). On the other hand, in the
40\,M$_{\odot}$ star case the SN explosion occurs at the centre of a
large hole in the molecular sheet, and the SN ejecta initially expand
almost unhindered through low density wind material (see
Fig. \ref{fig9}). A significant part of the SNR energy then appears to
be focused along the wind-blown cavity, and is carried away from the
molecular cloud. Nevertheless, we find that the molecular sheet is
still pushed back on Myr timescales and is eventually disrupted. Our
work clearly highlights the importance of capturing the effect of the
preceding stellar wind phase on the resulting SNR dynamics.
 
Turbulence resists the shaping effect of stellar winds. High levels of
turbulence ($Mach\sim$\,2-3) have been shown
\citep{offner15,kortgen16} to resist and dilute the effects of winds
and SN. However, questions have now arisen over the presence of such
turbulent conditions in dark molecular clouds. \cite{hacar16} claim that
the supersonic line-spreading taken to indicate supersonic turbulence
in the L1517 region of the Taurus cloud is in fact misleading and at most 
transonic large-scale velocities are present in the small-scale dark cores,
giving rise to supersonic lines through superposition of multiple transonic
line velocity components separated by supersonic velocity differences.
We generate transonic large-scale
velocities in our model as material flows and cools onto the
filamentary sheet-like cloud. Our results would indicate that
structured transonic velocities do not pose a barrier to the shaping
effects of stellar winds. \cite{offner15} also found that stellar
mass-loss rates must be greater than 10$^{-7}$\,M$_{\odot}$\,yr$^{-1}$
in order to reproduce shell properties.  In the 15\,M$_{\odot}$ star
case we have considered, mass-loss rates are less than this limit for
the entire MS evolution of the star, rising from
10$^{-8}$\,M$_{\odot}$\,yr$^{-1}$ to only
10$^{-7.4}$\,M$_{\odot}$\,yr$^{-1}$. On the other hand, the
40\,M$_{\odot}$ star always has a mass-loss rate above this limit -
during its MS evolution it loses mass at between 10$^{-6.4}$ and
10$^{-6}$\,M$_{\odot}$\,yr$^{-1}$. Our results, albeit testing only
these two cases, appear to be in agreement with the Offner \& Arce
result - the wind from the 15\,M$_{\odot}$ has little effect on its
parent cloud, restricted to a few pc radius in the corrugated sheet of
the cloud. In contrast, the wind from the 40\,M$_{\odot}$ star
evacuates a large tunnel-like cavity through the cloud, strongly
affecting the parent molecular cloud within a $\sim$20\,pc radius of
the star's location.

\subsection{Bubbles Interacting With Flattened Clouds and Bipolar HII regions}

Observations of wind-blown-bubbles often reveal that the surrounding
cold gas has a ring-like, rather than a spherical, morphology
\citep{beaumont10,deharveng10}. These observations detect little CO
emission towards the bubble centres, with the implication that the
molecular clouds which the bubbles are embedded in have thicknesses of
only a few parsecs\footnote{However, this interpretation remains
  controversial. \cite{anderson11} instead claim that the majority of
  bubble sources are three-dimensional structures.}. The current work
is highly relevant to such observations. For instance, Fig.~\ref{fig8}
shows that when a massive star is located in a relatively thin sheet
the bubble that forms initially has a bipolar structure and a ring of
swept-up material forms within the sheet. If the magnetic field in the
surrounding medium is perpendicular to the sheet the out-flowing wind
material is then directed along the field lines and flows
perpendicular to the molecular sheet.  Although we have not included  
photoionization in our simulations, we expect that  a bipolar HII  region 
would have  formed if we  had \citep[see,
e.g.,][]{bodenheimer79,deharveng15}.  A detailed comparison to such
observations is clearly warranted.

\section{Summary and conclusions}\label{conclusions}

In this work we have explored the effects of mechanical stellar wind and
supernova (SN) feedback on realistic molecular clouds in magnetic and
thermal pressure equilibrium. Our initial condition has been based on
the work of \cite{wareing16} in which a diffuse atomic cloud was
allowed to form structure through the action of the thermal
instability, under the influence of gravity and magnetic fields, but without injected turbulence. The
resulting structure is best described as a corrugated molecular sheet
surrounded by a diffuse atomic cloud. The molecular sheet appears
filamentary in projection. Once densities in the sheet had reached 100
cm$^{-3}$, a single massive star was introduced at the highest
density location after a further gravitational free-fall time of
$\sim$5\,Myrs. We considered two cases: formation of a 15\,M$_{\odot}$
star or a 40\,M$_{\odot}$ star. Their stellar winds (based on
realistic Geneva stellar evolution models) subsequently affect the parent
cloud, and at the end of each star's life a SN explosion is modelled.

In the 15\,M$_{\odot}$ star case the stellar wind has very little
effect on the molecular cloud, forming only a small bipolar
cavity. The SN is virtually oblivious to the wind cavity: instead, it
is primarily shaped by the presence of the molecular sheet and the
magnetic field which threads the cloud. The SN strongly affects the
sheet-like molecular cloud and its diffuse atomic surroundings. The
remnant sweeps up molecular material, creating a ring in the sheet. In
the diffuse cloud the remnant achieves a somewhat bipolar
morphology. The interaction between the remnant and the molecular
sheet results in material in the sheet being ablated into the SN
remnant. The ablated material is comparatively hot and dense and hence
likely to radiate strongly enough to be observable.
Additional simulations at higher resolution show that high
densities are achieved in the molecular sheet where ablation is taking
place, raising the possibility of SN-triggered star formation, and
also convince us that we are resolving the cooling in these
simulations accurately.

The SNR also temporarily raises the initial magnetic field strength by
a factor of 10 (in line with observations of molecular clouds that
show strong magnetic fields). Field strengths four times greater than
the initial field strength persist for hundreds of thousands of
years. At late time (6\,Myrs) the SN has largely dispersed the
molecular cloud into multiple fragments, with some 
evidence for the formation of further cold cloud material after the SN
caused material to become thermally unstable again. 
However, the domain size affects the simulations once
the SNR shock reaches the grid boundary, so interpretation of results
past this point must be treated with care.

In contrast, in the 40\,M$_{\odot}$ star case, the stellar wind has a
significant effect on the molecular cloud, forming a large tunnel-like
cavity through the centre of the corrugated-sheet-like molecular
cloud. In this case, the stellar wind also affects the
magnetic field, initially increasing field strengths by factors of
three or more, after which they return to close to their initial
values. At the end of the main sequence, the star enters the Luminous
Blue Variable (LBV) and then Wolf-Rayet (WR) phases of stellar
evolution. The result at the end of the WR phase is a dense shell of
mostly LBV material which has been swept up by the faster WR
wind. This structure is nearly spherical and has a diameter of
$\sim$25\,pc at the moment the star explodes as a SN.  Once the SNR
has swept up the LBV/WR shell, the forward shock rapidly progresses
along the tunnel-like cavity, directing significant energy away from
the sheet-like molecular cloud. The initial impact of the SN on the
molecular sheet is also reduced by the additional distance at which
the sheet lies from the explosion site.

It can be concluded from this work that SNe from the lower mass end of
the range of stars undergoing core-collapse have an equally important,
if not greater effect, on their parent molecular cloud than
higher-mass stars, on an individual star by star basis. The weakened
forward SN shock in the 40\,M$_{\odot}$ star case continues to expand
into the molecular cloud and diffuse cloud, but has less of an effect
on it as it cools into the radiative phase of its evolution, and its
shell breaks up through Rayleigh-Taylor instabilities after
500\,kyrs. Again, at late time, the molecular cloud is completely disrupted.
Cold cloud material still exists in the simulations, 
indicating cloud remnants are still present, but there is no structural
resemblance to the corrugated sheet-like nature of the initial condition. 
There is evidence that both stars cause stable material to return to the 
thermally unstable phase after the SN and in the 40\,M$_\odot$ star case, 
the amount of cold, dense material shows a sharp increase 3\,Myrs post-SN.
However, since boundary effects may be influencing our results at this
time further numerical work is required to consider this late part of the cloud evolution 
more accurately.

Similarities exist between the stellar wind bubble in the
40\,M$_{\odot}$ star case and the rings and bubbles seen in some
observations \citep{beaumont10,deharveng10}. Our work is also relevant
to understanding structures created by a cluster of young massive
stars, such as the Rosette Nebula, and even to understanding bipolar
HII regions \citep[e.g.][]{deharveng15}. Further work is required to
explore these similarities and to directly compare detailed models to
the wealth of relevant observational data.

It should also be noted in this work that we have taken a single
initial condition, realistic but formed from the action of the thermal
instability. Different initial conditions could lead to considerably
different conclusions, the key parameters being the amount of mass and
degree of inhomogeneity of the parent molecular cloud, and the
pressure, density and magnetic field strength of the
environment. Molecular cloud masses in the Milky Way reach
10$^{5-6}$\,M$_{\odot}$, considerably more than the
17,000\,M$_{\odot}$ of material in the cloud investigated here. The
nature and distribution of this material is key to how the stellar
winds and SN affect the molecular cloud. In future work we will
explore further how these parameters affect the resistance of
molecular clouds to winds, supernovae, and ionising radiation.

\section*{Acknowledgments}

This work was supported by the Science \& Technology Facilities
Council [Research Grant ST/L000628/1]. The calculations for this paper
were performed on the DiRAC Facility jointly funded by STFC, the Large
Facilities Capital Fund of BIS and the University of Leeds and on
other HPC facilities at the University of Leeds. These facilities are
hosted and enabled through the ARC HPC resources and support team at
the University of Leeds (A. Real, M.Dixon, M. Wallis, M. Callaghan \&
J. Leng), to whom we extend our grateful thanks.  
We acknowledge useful discussions with T. W. Hartquist
and S. Van Loo and detailed review from the anonymous reviewer which
improved this work. We also extend our thanks to S. Van Loo for the provision 
of analysis routines to produce PDFs. Data for the all figures in this paper 
is available from http://doi.org/10.5518/114. 
VisIt is supported
by the Department of Energy with funding from the Advanced Simulation
and Computing Program and the Scientific Discovery through Advanced
Computing Program.

\appendix

\section{Line profiles}

\begin{figure*}
\centering
\includegraphics[width=175mm]{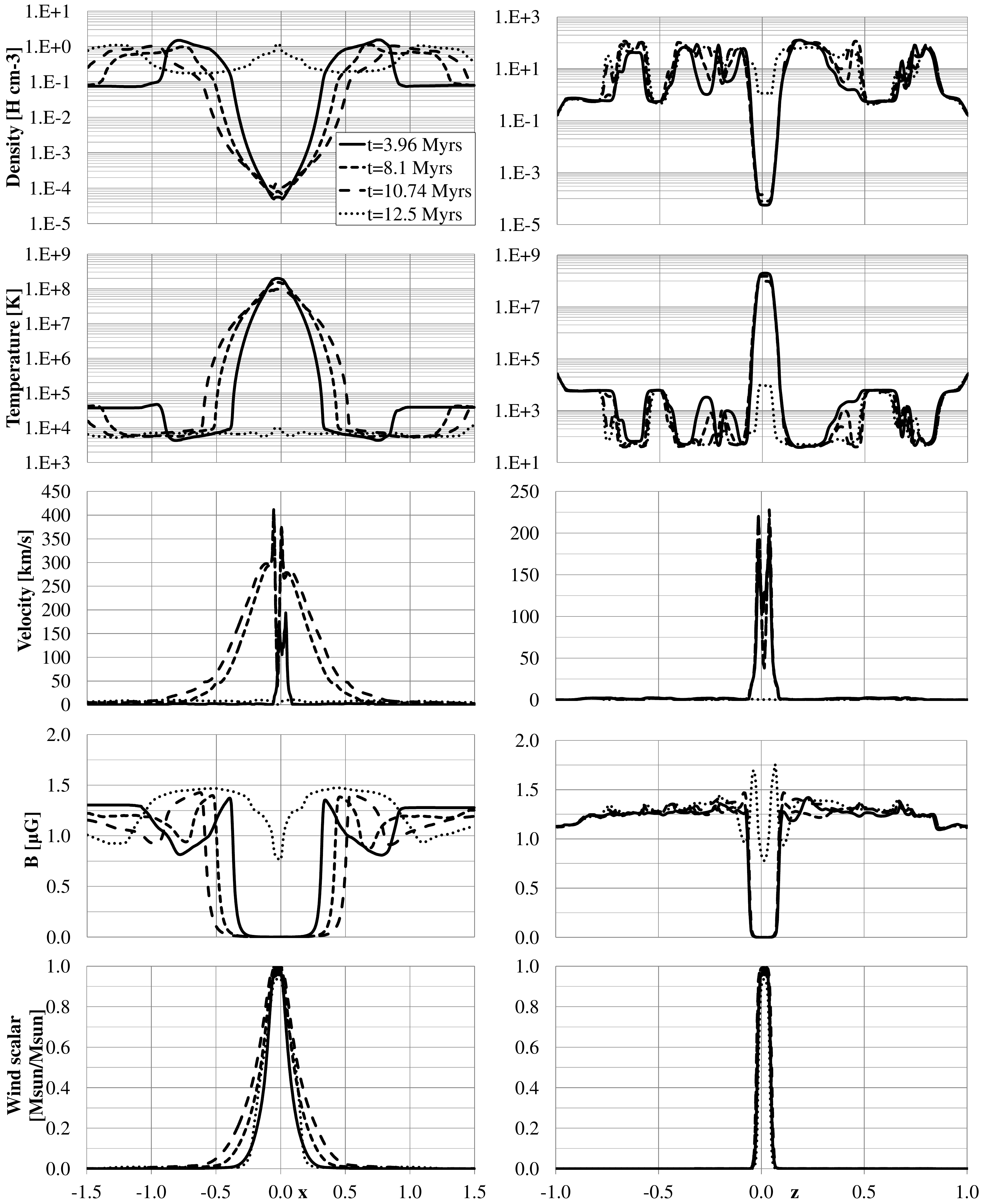}
\caption{Line profiles of the wind-cloud structure
    shown in Figure \ref{fig3} for the 15 M$_\odot$ star simulation.
Left column: profiles along the magnetic field in the $x$ direction at the location of the
central star ($y$,$z$) = (0.0, 0.0125).
Right column: profiles across the magnetic field in the $z$ direction and across the 
corrugated sheet of the molecular cloud at the location of the central star 
($x$,$y$) = (-0.025, 0.0).} 
\label{figA1}
\end{figure*}

\begin{figure*}
\centering
\includegraphics[width=175mm]{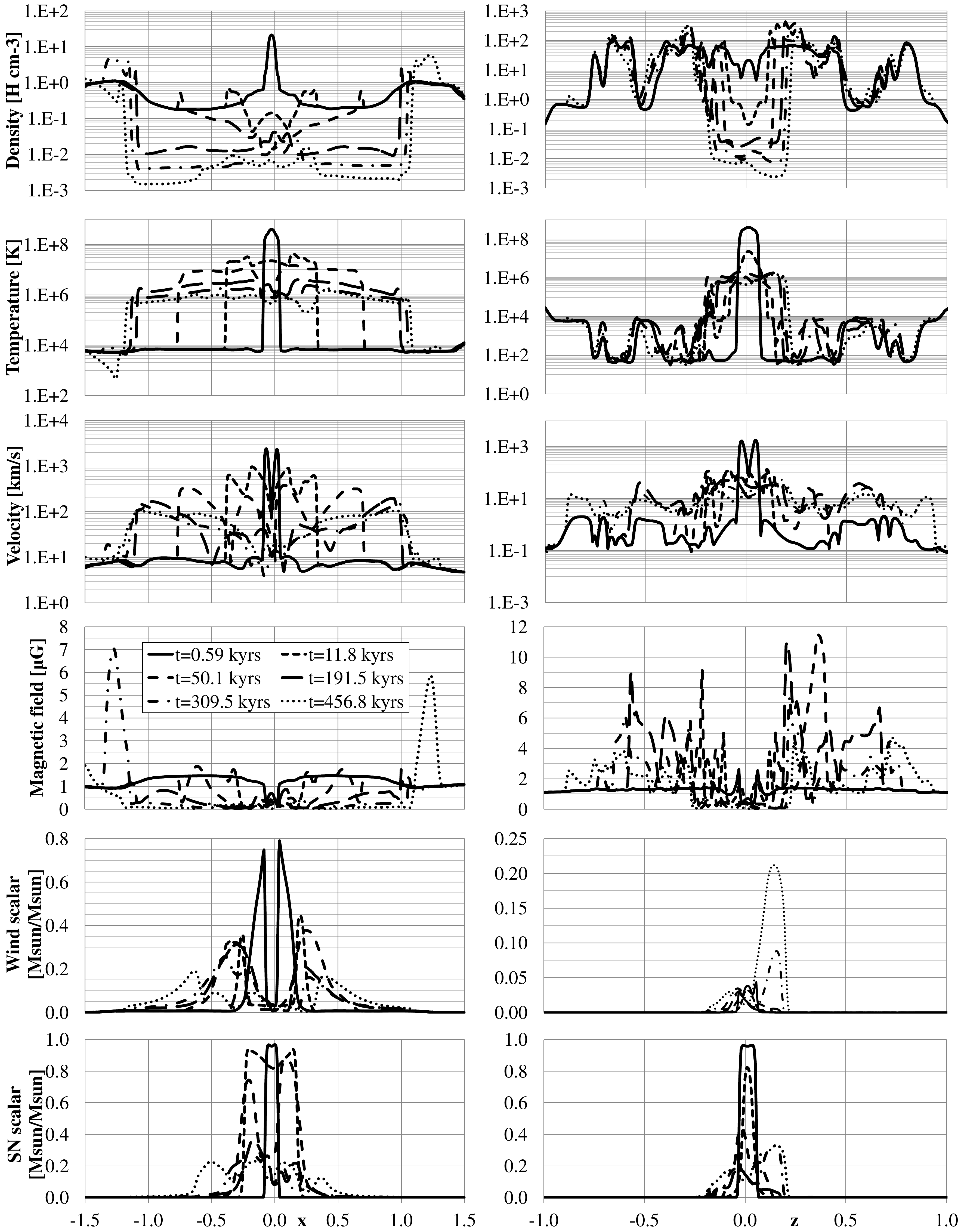}
\caption{Line profiles of the SN-wind-cloud structure
    shown in Figure \ref{fig4} for the 15 M$_\odot$ star simulation.
Left column: profiles along the magnetic field in the $x$ direction at the location of the
central star ($y$,$z$) = (0.0, 0.0125).
Right column: profiles across the magnetic field in the $z$ direction and across the 
corrugated sheet of the molecular cloud at the location of the central star 
($x$,$y$) = (-0.025, 0.0).} 
\label{figA2}
\end{figure*}

\begin{figure*}
\centering
\includegraphics[width=175mm]{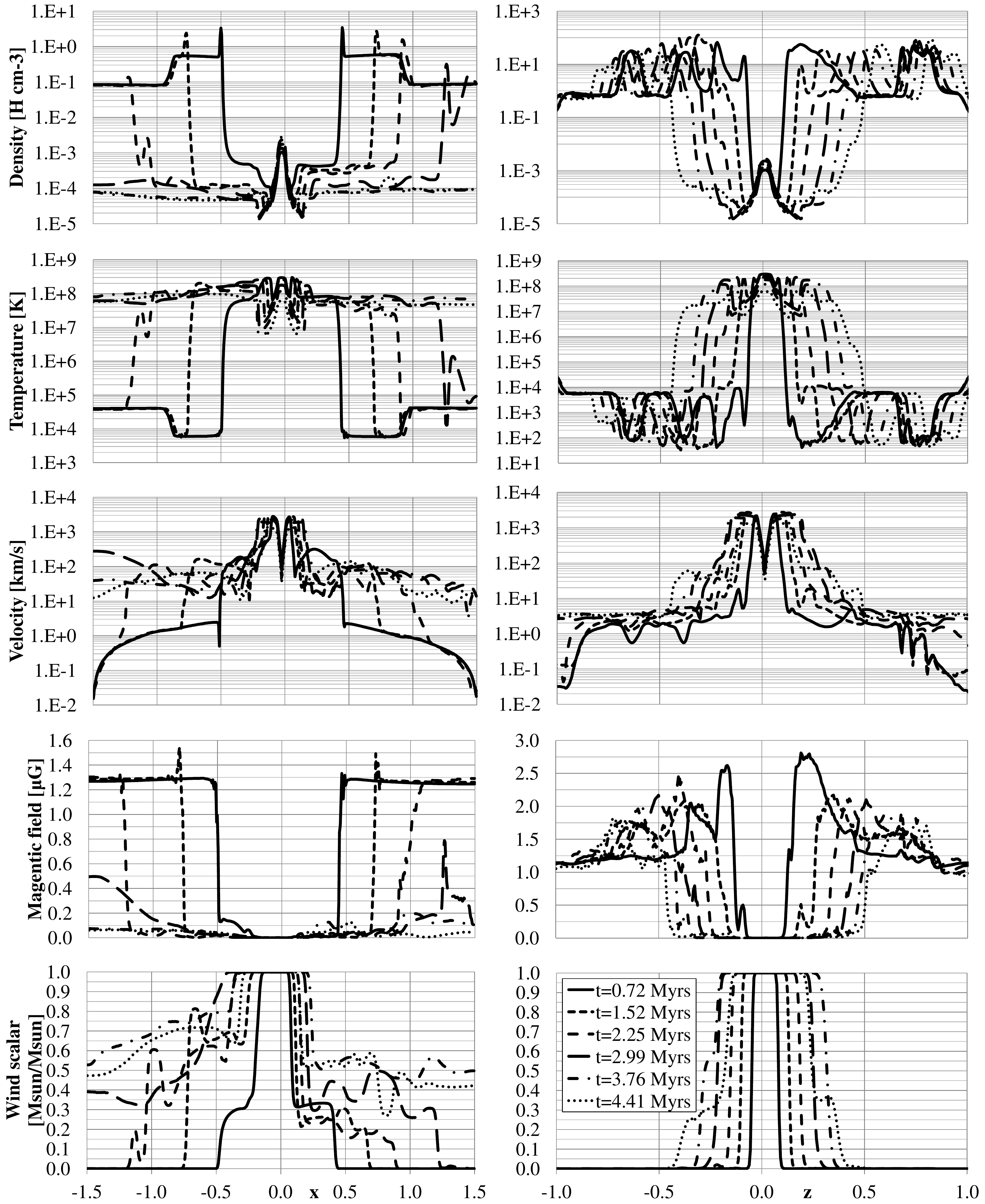}
\caption{Line profiles of the wind-cloud structure
    shown in Figure \ref{fig8} for the 40 M$_\odot$ star simulation.
Left column: profiles along the magnetic field in the $x$ direction at the location of the
central star ($y$,$z$) = (0.0, 0.0125).
Right column: profiles across the magnetic field in the $z$ direction and across the 
corrugated sheet of the molecular cloud at the location of the central star 
($x$,$y$) = (-0.025, 0.0).}
\label{figA3}
\end{figure*}

\begin{figure*}
\centering
\includegraphics[width=175mm]{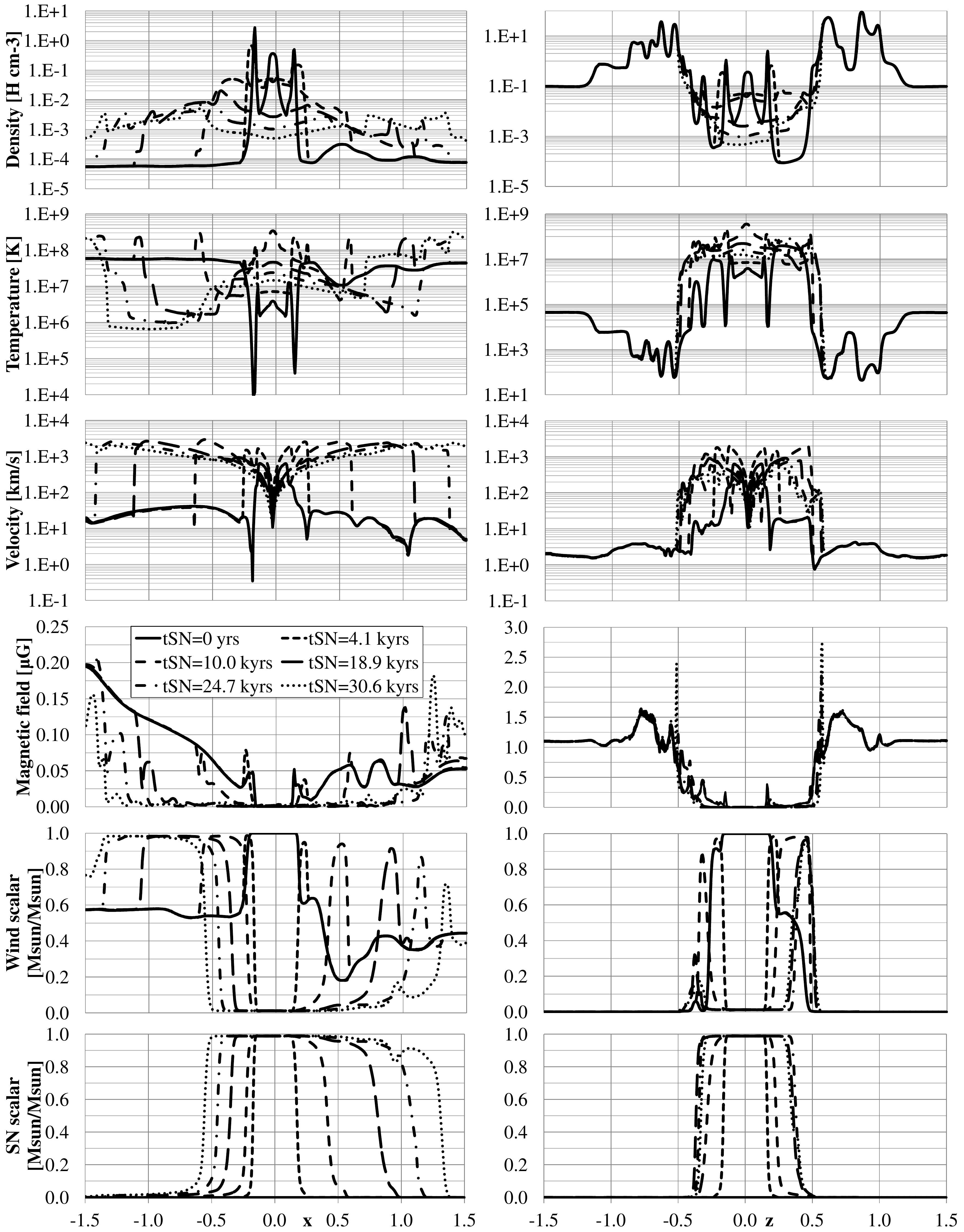}
\caption{Line profiles of the early-time
    SN-wind-cloud structure shown in Figure \ref{fig9} for the 40
    M$_\odot$ star simulation.
Left column: profiles along the magnetic field in the $x$ direction at the location of the
central star ($y$,$z$) = (0.0, 0.0125).
Right column: profiles across the magnetic field in the $z$ direction and across the 
corrugated sheet of the molecular cloud at the location of the central star 
($x$,$y$) = (-0.025, 0.0).}
\label{figA4}
\end{figure*}

\begin{figure*}
\centering
\includegraphics[width=175mm]{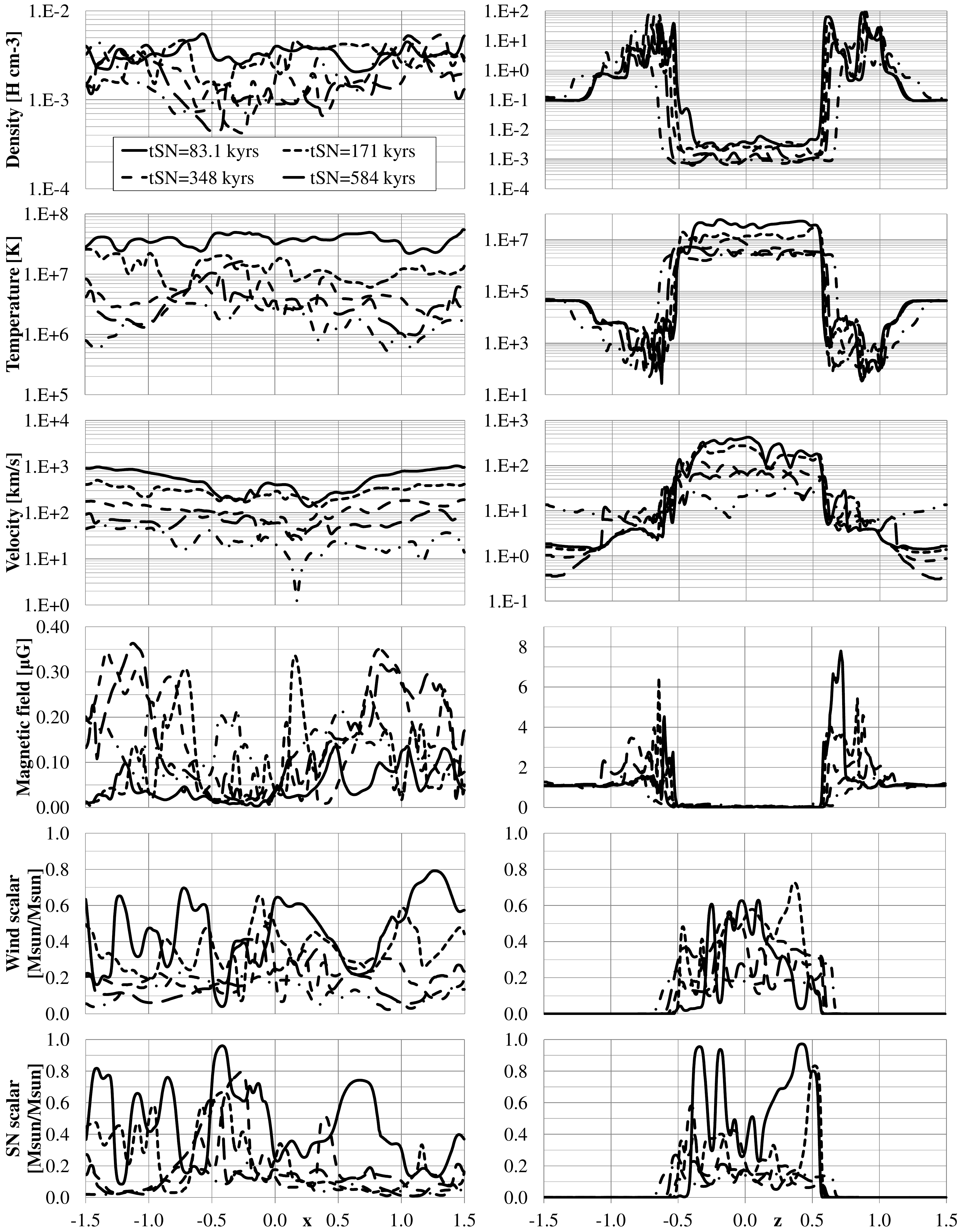}
\caption{Line profiles of the SN-wind-cloud structure
    shown in Figure \ref{fig10} for the 40 M$_\odot$ star simulation.
Left column: profiles along the magnetic field in the $x$ direction at the location of the
central star ($y$,$z$) = (0.0, 0.0125).
Right column: profiles across the magnetic field in the $z$ direction and across the 
corrugated sheet of the molecular cloud at the location of the central star 
($x$,$y$) = (-0.025, 0.0).}
\label{figA5}
\end{figure*}

\label{lastpage}

\end{document}